\documentclass[aps,prd,twocolumn,nofootinbib,superscriptaddress]{revtex4-1}

\usepackage{amsfonts}
\usepackage{amsmath}
\usepackage{amssymb}
\usepackage{bm}
\usepackage{dcolumn}
\usepackage{epsfig}
\usepackage{graphics}
\usepackage[latin1]{inputenc}
\usepackage{latexsym}
\usepackage{rotating}
\usepackage[colorlinks=true]{hyperref}

\hypersetup{%
   citecolor=black
}

\usepackage[usenames]{color}
\usepackage{float}
\usepackage{ucs}
\usepackage{xspace} 
\usepackage{mathrsfs}
\usepackage{subfigure}
\usepackage{enumitem}
\usepackage{tabularx}
\usepackage{booktabs}
\usepackage{array}
\usepackage[normalem]{ulem}

\def\be{\begin{equation}}
\def\ee{\end{equation}}
\newcommand{\bea}{\begin{eqnarray}}
\newcommand{\eea}{\end{eqnarray}}

\newcommand{\dd}[1]{\mathrm{d}#1\,}
\DeclareMathOperator{\sgn}{sgn}
\DeclareMathOperator{\erf}{erf}

\newcolumntype{Y}{>{\centering\arraybackslash}X}

\DeclareMathOperator{\ratqm}{\tilde{q}}

\begin{document}

\title{Hidden-Sector Modifications to Gravitational Waves From Binary Inspirals}

\author{Stephon Alexander}
\email{stephon\_alexander@brown.edu}
\affiliation{Department of Physics, Brown University, Providence, RI, 02906}

\author{Evan McDonough}
\email{evan\_mcdonough@brown.edu}
\affiliation{Department of Physics, Brown University, Providence, RI, 02906}

\author{Robert Sims}
\email{robert\_sims@brown.edu}
\affiliation{Department of Physics, Brown University, Providence, RI, 02906}

\author{Nicol\'as Yunes}
\email{nicolas.yunes@montana.edu}
\affiliation{eXtreme Gravity Institute, Department of Physics, Montana State University, Bozeman, MT 59717, USA.}

\begin{abstract}

Gravitational wave astronomy has placed strong constraints on fundamental physics, and there is every expectation that future observations will continue to do so. In this work we quantify this expectation for future binary merger observations to constrain hidden sectors, such as scalar-tensor gravity or dark matter, which induce a Yukawa-type modification to the gravitational potential. We explicitly compute the gravitational waveform, and perform a Fisher information matrix analysis to estimate the sensitivity of next generation gravitational wave detectors to these modifications. We find an optimal sensitivity to the Yukawa interaction strength of $10^{-5}$ and to the associated dipole emission parameter of $10^{-7}$, with the best constraints arising from the Einstein Telescope. When applied to a minimal model of dark matter, this provides an exquisite probe of dark matter accumulation by neutron stars, and for sub-TeV dark matter gravitational waves are able to detect mass fractions $m_{DM}/m_{NS}$ less then 1 part in $10^{15}$.

\end{abstract}

\date{\today}

\maketitle

\section{Introduction}

With the observation of black hole binary mergers \cite{Abbott:2017gyy,Abbott:2017vtc,Abbott:2016nmj,Abbott:2016blz,Abbott:2017oio} and a neutron star binary merger \cite{TheLIGOScientific:2017qsa}, gravitational wave astronomy is rapidly emerging as a powerful probe of fundamental physics \cite{Chamberlain:2017fjl}. These observations provide an exquisite confirmation of General Relativity in the extreme gravity regime, placing severe constraints on modifications to gravity \cite{Yunes:2016jcc}, and ruling out large classes of dark energy models invoked to explain the current acceleration of the universe \cite{Ezquiaga:2017ekz, Creminelli:2017sry,Sakstein:2017xjx}.  

Future gravitational wave observations should also be sensitive probes of other modifications to General Relativity, such as large extra-dimensions \cite{Visinelli:2017bny,Pardo:2018ipy}, time-varying fundamental constants \cite{Amendola:2017ovw}, parity violation \cite{ay1,ay2,ay3}, Lorentz violation \cite{Yagi:2013qpa,Yagi:2013ava}, scalar-tensor gravity \cite{Yunes:2011aa,Damour:1993hw,Damour:1996ke,Mirshekari:2013vb,Lombriser:2016yzn}, and dark matter \cite{Goldman:1989nd,Kouvaris:2007ay,Kouvaris:2010vv,deLavallaz:2010wp,1012.2039,1103.5472,1201.2400,1301.0036,1301.6811,1310.3509,Bramante:2017xlb,Bramante:2017ulk,Zheng:2014fya, Sandin:2008db,Croon:2017zcu,Ellis:2017jgp,Kopp:2018jom} (see also \cite{Bird:2016dcv}). These modifications can by probed by precise interferometer measurements of the gravitational waves emitted by compact binary mergers, though to do so requires building analytic templates of the modified waves and a detailed statistical analysis. In this work, we undertake precisely this task, focusing on modifications that induce a Yukawa-type modification to the gravitational potential, and with a particular focus on dark matter.

This latter possibility, i.e.~gravitational waves as a probe of dark matter, is particularly relevant given the current state of dark matter observations. Indeed, the only observational evidence for dark matter is gravitational, e.g.~the peaks of the cosmic microwave background \cite{Aghanim:2018eyx}, galactic rotation curves \cite{Rubin:1970zza,Rubin:1980zd}, and the Bullet Cluster \cite{Clowe:2006eq}. This motivates the study of dark matter with an eye towards the new observational window into gravitational physics: gravitational waves from binary inspirals. Parallel to this, the connection of \emph{primordial} gravitational waves to dark matter has been proposed in \cite{Alexander:2018fjp}; these gravitational waves are, however, most readily observed via the polarization of the cosmic microwave background \cite{Guzzetti:2016mkm}.

The connection of binary mergers to dark matter arises through the possibility that dark matter is gravitationally bound inside of neutron stars \cite{Goldman:1989nd,Kouvaris:2007ay,Kouvaris:2010vv,deLavallaz:2010wp,1012.2039,1103.5472,1201.2400,1301.0036,1301.6811,1310.3509,Bramante:2017xlb,Bramante:2017ulk,Zheng:2014fya, Sandin:2008db,Raj:2017wrv}. If the dark sector includes a light force mediator, then this naturally leads to an additional force between neutron stars, similar to that experienced by compact objects in scalar-tensor gravity, where the role of accumulated mass is played by a scalar field-dependent modulation of the inertial mass. This additional force modifies the gravitational wave signal from neutron star binary mergers, which can potentially probe the underlying dark matter model \cite{Croon:2017zcu,Ellis:2017jgp,Nelson:2018xtr, Kopp:2018jom} (see also \cite{Huang:2018pbu}). We emphasize that this conclusion is \emph{completely general} and it does not depend on a specific dark matter model.  

The amount of dark matter inside neutron stars is subject to considerable theoretical uncertainty, since this does depend, not only on the dark matter model, but also on the formation and entire lifetime of the neutron star. Estimates of the fraction of the neutron star mass in dark matter range from a few percent \cite{Ellis:2018bkr} to one part in  $10^{15}$ \cite{Kopp:2018jom}. Remarkably, in this work we find that gravitational wave observations can probe dark matter even at mass fractions below the latter estimate. Independent of the underlying model, however, be it dark matter or a scalar-tensor theory, the modification to the binary inspiral is induced by a simple Yukawa correction to the gravitational potential, of the form $V_\text{Yuk}/V_\text{grav} \sim \alpha e^{-r/\lambda}$, where $\alpha$ is a constant parametrizing the relative strength of the interaction and $\lambda$ is the length scale of the interaction, which corresponds to a dark force mediator mass $m_v$ of $\lambda/{\rm km} \equiv 9.73 \times 10^{-11} ( {\rm eV}/m_v)$. 

For the minimal dark matter model we consider, $\alpha<0$ and the force is repulsive.  At a given fixed orbital separation $r$, this results in a decrease in both the orbital frequency and the total energy of the system, leading in turn to a decrease in the power emitted in gravitational waves (again at a given fixed $r$). In addition to this effect, and irrespective of the particular dark matter model considered, when the orbital angular frequency of the binary exceeds $1/\lambda$, an associated dipole emission is activated, which forces the binary to inspiral faster. This dipole emission will be dominant over the quadrupole emission of General Relativity, dominating the balance law and the chirping rate at large separation. 

\begin{figure}
\begin{center}
\includegraphics[width=.5 \textwidth]{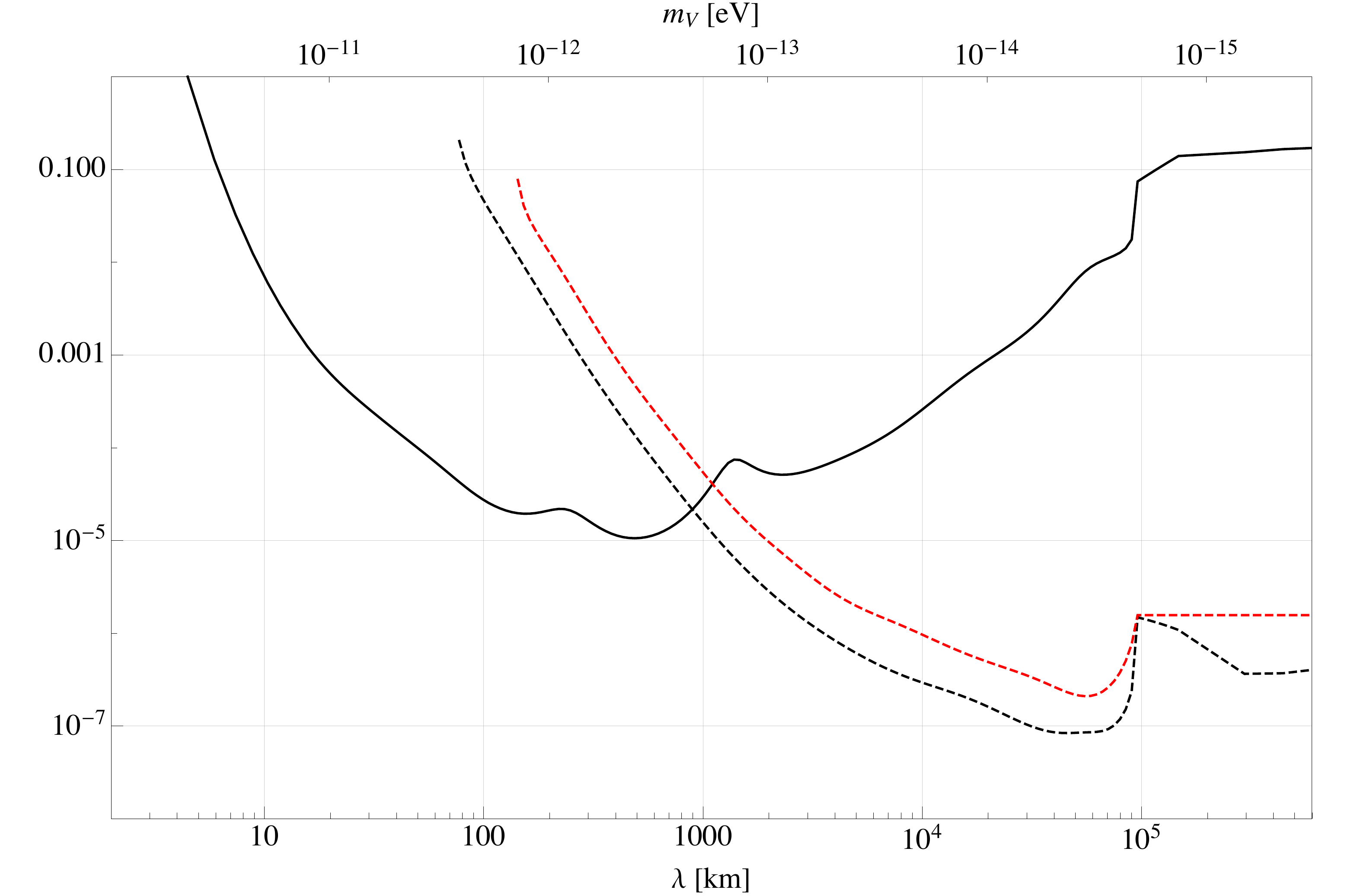}
\caption{ Projected sensitivity of the Einstein Telescope ET-B \cite{Hild:2009ns, ET} to the relative Yukawa interaction strength $\alpha$ (black, solid), and the dipole emission parameter $\gamma$ for a neutron star binary (black, dashed) and a mixed black hole-neutron star binary (red,dashed). Dipole emission occurs only for $\lambda$ above a mass-dependent lower bound.}
\label{Fig:preview}
\end{center}
\vspace{0cm}
\end{figure}

In this work we explicitly compute analytic expressions for the Fourier space gravitational wave amplitude and phase in the stationary phase approximation, given the above form of the correction to the potential. We find that the induced dipole emission manifests itself as a -1 post-Newtonian (PN) correction to the waveform, while the Yukawa-type corrections in general do not decompose into a PN expansion. In the limit of a light dark photon $\lambda \gtrsim \mathcal{O}(10^3) \, {\rm km}$, the Yukawa corrections can be formulated as a (convergent) series of negative PN corrections,  while in the opposite limit, $\lambda \lesssim 5 {\rm km}$, the corrections are exponentially suppressed and become completely degenerate with the General Relativity waveform. 

We perform a Fisher information matrix analysis and compute the projected sensitivity to the above dark sector modifications  of second and third generation ground-based gravitational wave interferometers: advanced LIGO \cite{Aasi:2013wya}, A+/A++\cite{LIGOwhitepaper}, Voyager \cite{LIGOwhitepaper}, VRT \cite{Adhikari:2013kya,LIGOwhitepaper}, Cosmic Explorer \cite{LIGOwhitepaper} and the Einstein Telescope \cite{Hild:2009ns, ET}.  This approach to statistical inference has become a standard tool in gravitational wave physics (see e.g.~\cite{Chamberlain:2017fjl}), provided the signal is loud enough and the noise is stationary and Gaussian, as expected in third generation detectors. Such a tool is highly computational efficient, allowing for an expedient search for areas of new physics `where the light shines brightest' with gravitational waves. In this work we find that the light shines very brightly on dark sectors, and in Figure \ref{Fig:preview} we provide a preview of our main results. This figure shows that a single observation with the Einstein Telescope would suffice to rule out a large sector of parameter space:  values of the Yukawa interaction strength $\alpha$ above the solid line, and values of the dipole emission parameter $\gamma$ above the dashed lines.  For a particular model, one can use black-hole superradiance \cite{Cardoso:2018tly}, in tandem with binary insprial gravitational wave systems, to probe particular length scales $\lambda$.

Our work extends previous analyses (e.g.~\cite{Sagunski:2017nzb,Cardoso:2016olt,Croon:2017zcu,Ellis:2017jgp,Nelson:2018xtr, Kopp:2018jom, Huang:2018pbu}\footnote{Note also \cite{Battye:2018ssx}, which takes a different approach from this work.}) to include projected constraints with a wider range of $\lambda$ and with second and third generation detectors, a careful accounting of projected constraints on dipole emission, and explicit expressions for the inspiral waveform that carefully include the effect of Yukawa and dipole modifications, which could be used in the future against real data. We find that the constraints on the Yukawa interaction strength $\alpha$ can be as good as $\mathcal{O}(10^{-5})$, with the best constraints coming from the Einstein Telescope, while constraints on the dipole emission parameter $\gamma$ (defined in Eq.~\eqref{gammadef}) can be as good as $\mathcal{O}(10^{-7})$, with the best constraints again coming from the Einstein Telescope. From this we conclude that gravitational waves are indeed a powerful probe of dark sectors.

The structure of this paper is as follows. Section~\ref{Sec:Model} outlines an example of a dark model which realizes the discussed modifications to the gravitational potential. Section~\ref{sec:mod-GWs} discusses general features of the modified inspiral, and Sec.~\ref{Sec:Waveform} explicitly computes the gravitational waveform. Section~\ref{Sec:MethodofConstraints} performs a Fisher analysis on the waveform, with our main results presented in Figures \ref{fig:gammaplot} and \ref{fig:alphaplotNSNS}. We conclude in Section~\ref{sec:conclusions} with a summary of our results and a discussion of directions for future work.

\section{Dark Matter Model}
\label{Sec:Model}

As we discussed in the Introduction, our work is largely independent of the dark-matter model. However, for the sake of concreteness, we here provide a specific example which realizes this scenario. We emphasize, nonetheless, that the results presented in this paper are generic and not dependent of the specific features a particular dark matter model. 

Consider then a model of asymmetric dark matter (for a review see e.g.~\cite{Petraki:2013wwa}) coupled to an Abelian gauge field $V^\mu$ (the `dark photon'), as has been considered previously in e.g.~\cite{Petraki:2014uza,Kopp:2018jom}. The dark sector Lagrangian\footnote{With the exception of the Lagrangian, we will use geometric units throughout our analysis.} is given as
\begin{equation}
\mathcal{L}_\text{DS} = -\frac{1}{4}V_{\mu\nu}V^{\mu\nu}+\frac{1}{2}m_v^2 V_\mu V^\mu + \bar{\chi}\left(i\gamma^\mu D_\mu - m_\chi\right)\chi, \label{EQ:DSLagr}
\end{equation}
where $D_\mu = \nabla_\mu + i g V_\mu$ is the gauge covariant derivative, $V_{\mu\nu}$ is the dark photon field strength tensor, and the fermion $\chi$ has dark charge $g$ and mass $m_{\chi}$.  The dark photon mass $m_v$ can arise through a Higgs or Stueckelberg mechanism, but such completions of the theory will produce negligible effects on our analysis.  Further, one can generalize this Lagrangian to non-Abelian gauge fields, but the lightest, massive gauge field will produce the most noticeable change in gravitational waves.

The range of dark photon masses that can be probed by gravitational waves are extremely light, $m_{v} \lesssim 10^{-10}$ eV, and gauge invariance is approximately conserved. This implies that a charge asymmetry for $\chi$ must be balanced by an opposite charge asymmetry for a second fermion, analogous to standard electromagnetism and the protons and electrons in our current universe. This opens up the possibility that some fraction of the dark matter will form neutral bound states, the precise value of which depends sensitively on the value of the dark photon's fine structure constant \cite{Petraki:2014uza}.  Capture of these bounds states in compact objects will contribute to the dark dipole moment at lowest order, however we only consider the corrections due to a dark monopole moment here.

In order to produce a nonzero dark monopole moment, a net charge asymmetry will be required for neutron stars.  Neutron stars can receive dark matter from two sources: (1) dark matter accreted from the surrounding halo, and (2) dark matter contained in the progenitor. The former has been studied in detail in \cite{Goldman:1989nd,Kouvaris:2007ay, Kouvaris:2010vv, deLavallaz:2010wp, 1012.2039, 1103.5472, 1201.2400, 1301.0036, 1301.6811, 1310.3509, Bramante:2017xlb, Bramante:2017ulk, Ellis:2018bkr}, and most recently in \cite{Kopp:2018jom}. The latter has been argued to open up the possibility of anywhere from a few percent to an $\mathcal{O}(1)$ fraction of the mass of a neutron star to be dark matter, a so-called `Admixture Neutron Star' \cite{Leung:2011zz, Rezaei:2018cuk}.

If one considers only the accretion of dark matter by neutron stars, the number of  dark matter particles\footnote{Considering a neutron star with mass $1.44 M_\odot$ and radius $10.6$ km.} with $m_{\chi}\gtrsim 1$ GeV that are captured can be estimated as \cite{1103.5472,Nelson:2018xtr},
\begin{align}
N_{\chi} \simeq 2.3 &\times 10^{44} \left(\frac{100 \text{ GeV}}{m_{\chi}}\right) \left(\frac{\rho_{\chi}}{10^3 \text{ GeV/cm}^3}\right)\nonumber\\
\times &\left(\frac{\sigma_{B}}{2.1 \times 10^{-45} \text{ cm}^2}\right) \left( \frac{t_{NS}}{10^{10} \text{ yr}}\right) , \label{EQ:NumChi}
\end{align}
where $t_{NS}$ is the age of the neutron star, and $\sigma_B$ is the lesser of the DM-neutron elastic scattering cross section $\sigma_n$ and the effective geometric scattering cross section.  For lighter dark matter, the number of accreted particles is independent of the dark matter mass \cite{1103.5472}.  Therefore, if there is a mass difference between the two dark matter fermions, and at least one is lighter than a GeV, a net charge can accumulate and the accretion is predominantly into the heavier $\chi$ fermions.

From the number of dark matter particles accreted, the fraction of the neutron star mass in the form of dark matter $f_{DM} = N_\chi m_\chi / m_\text{NS}$ can be approximated to $f_{DM} \simeq 10^{-11}$ assuming standard parameters. Similar estimates have been made in the literature, with varying levels of precision. The most recent estimate is given by \cite{Kopp:2018jom}, which gives a more conservative bound of $f_{DM} \lesssim 10^{-15}$.

As we will show in Section \ref{Sec:Constraints}, gravitational waves can still probe these small charge accumulations in compact objects.  The relative strength of the dark photon's Yukawa interaction compared to gravity can compensate for the small dark matter fraction.  This relative strength can be approximated as
\begin{equation}
\alpha \approx 1.18\times 10^{33} g^2 f_{DM} ^2\left(\frac{100 \text{ GeV}}{m_\chi}\right)^2. \label{EQ:AlphaFrac}
\end{equation}
Even using the conservative bound $f\sim 10^{-15}$, we see that the dark Yukawa interaction can remain relatively strong for weakly coupled ($g\ll 1$) dark fermions.

\section{Modifications to Gravitational Wave Physics of Binary Inspirals}
\label{sec:mod-GWs}

Given our simple dark matter model, we now consider the dynamic effects that manifest with a net dark charge on the binary system.  During the early stages of the inspiral, the binary constituents are treated as point masses/charges.  In this regime, the interaction between the two compact object via the dark photon can be approximated as a tree-level scattering.  This interaction will manifest as a Yukawa correction to the potential energy of the binary system, given by
\begin{equation}
V_\text{Yuk}(r) = \alpha\frac{m^2\eta}{r} e^{-r/\lambda},\label{EQ:TotPot}
\end{equation}
where $\lambda = m_v^{-1}$ is the length scale of the Yukawa interaction, $m=m_1+m_2$ is the total mass of the binary, $\eta = m_1 m_2 /m^2$ is the symmetric mass ratio, $r$ is the orbital separation, and the relative strength of the Yukawa potential $\alpha$, from Eq.~(\ref{EQ:AlphaFrac}), can be defined in terms of the neutron star properties as
\begin{equation}
\alpha = \frac{q_1 q_2}{m_1 m_2} = \ratqm_1 \ratqm_2\label{EQ:StrengthDef}
\end{equation}
where $\ratqm_i = q_i/m_i$ is the dark charge to mass ratio of each star.  For the asymmetric dark matter model we consider, both compact objects should acquire the same sign of net dark charge, thus we work in the regime where $\alpha >0$\footnote{For a scalar mediator, the argument presented would give $\alpha < 0$, i.e. an attractive interaction.}.

This modification to the potential ultimately leads to a violation of Kepler's laws which will be functionally distinct from General Relativity corrections.  For (nearly) circular orbits, the modification will manifest as\footnote{Our formula differs from \cite{Croon:2017zcu} by an additional factor of the Yukawa term, agreeing with \cite{Kopp:2018jom}.}
\begin{equation}
\omega^2 = \frac{1}{m\eta r}\frac{dV}{dr} = \frac{m}{r^3}\left[1-\alpha\left(1+\frac{r}{\lambda}\right)e^{-r/\lambda}\right].\label{EQ:OrbFreq}
\end{equation}
Furthermore, the potential is no longer a power law, hence the Virial theorem takes a more complicated form when evaluating the total energy of the binary.  The latter can be calculated as
\begin{equation}
E_\text{tot} = -\frac{m^2\eta}{2r}\left[1-\alpha \left(1-\frac{r}{\lambda}\right)e^{-r/\lambda}\right]. \label{EQ:Energy}
\end{equation}
The repulsive Yukawa potential results in both a decrease in the orbital frequency and magnitude of the total energy of the system at a given orbital separation.

These kinematic variables dictate the rate at which energy is radiated away from the system in the form of gravitational radiation.  The power emitted in the form of gravitational radiation can be computed from the quadrupole moment as
\begin{align}
P_\text{GW} &= \frac{D_L^2}{32\pi}\int{\dd  \Omega}\langle\dot{h}_{ij}^{TT}\dot{h}_{TT}^{ij}\rangle\nonumber\\
&= \frac{32}{5}\eta^2 m^2 \omega^6 r^4 = \frac{32}{5}\eta^2 v^{10},
\end{align}
where the dot represents a time derivative, $D_L$ is the luminosity distance, and $v = \omega r$ is the orbital velocity for a quasi-circular orbit.

When gravitational waves are the only form of emitted radiation, the balance law, $P_\text{GW} = -\frac{d}{dt}E_\text{tot}$, can be used to find the rate at which the orbital separation decreases as
\begin{equation}
\frac{d r}{d t} = - \frac{64\eta}{5}\left(\frac{m}{r}\right)^3\frac{\left[1-\alpha \left(1+\frac{r}{\lambda}\right)e^{-r/\lambda}\right]^3}{1-\alpha (1+\frac{r}{\lambda}-\frac{r^2}{\lambda^2})e^{-r/\lambda}}. \label{EQ:DiffRFull}
\end{equation}
While the power emitted in gravitational radiation is reduced due to the repulsive Yukawa interaction, this need not translate into a longer coalescence time than the Newtonian/General Relativity predictions.  Instead, the decrease in energy of the system in Eq.~(\ref{EQ:Energy}) can overcompensate for this decrease in radiation leading to a quicker inspiral phase.

\subsection{Dark dipole radiation}
\label{Sec:DarkRadMod}

Up to this point, we have ignored the on-shell emission of the dark photon due to the orbital motion of the charged compact objects.  However, this dipole radiation introduces an important, and potentially dominant, source of energy dissipation to the binary system.  The effect of dipole radiation on the binary dynamics has been studied in \cite{Croon:2017zcu,Krause:1994ar}.  In our context, the additional power radiated is given by\footnote{Relative to \cite{Croon:2017zcu}, we include an additional factor of two for the vector mode dipole radiation, consistent with the results of \cite{Krause:1994ar}.}
\begin{align}
P_\text{dark} &= \frac{2}{3}\gamma\eta^2m^2\omega^4r^2\left(1+\frac{1}{2(\lambda\omega)^2}\right)\left(1-\frac{1}{\lambda\omega}\right)^{\frac{1}{2}},\nonumber\\
&= \frac{2}{3}\gamma\eta^2 v^8\left(1+\frac{1}{2(\lambda\omega)^2}\right)\left(1-\frac{1}{\lambda\omega}\right)^{\frac{1}{2}},\label{EQ:PDark}
\end{align}
where
\be
\label{gammadef}
\gamma \equiv (\ratqm_1-\ratqm_2)^2
\ee
is the squared difference between the charge-to-mass ratios of the binary stars. Clearly, the effects of dipole radiation will only manifest when the dark matter mass fraction of the compact objects differ.

The other two functions of $\lambda \omega$ can be approximated as the Heaviside step function $\theta(\lambda\omega-1)$, but note that the functional form is not actually of Heaviside form; Appendix \ref{App:DipoleRadCorrections} computes the corrections between the above functional form and the Heaviside approximation.  The argument of the step function determines the activation of dipole radiation.  This relation can be written in terms of the Yukawa length scale $\lambda$ and the gravitational wave frequency $f$ as
\begin{equation}
\lambda \geq 9.5\times 10^3\text{ km}\left(\frac{10\text{ Hz}}{f}\right).\label{EQ:DipActive}
\end{equation}
For dipole radiation to be active, the Yukawa interaction must have a length scale much larger than the orbital separation of the binary.  As we will see in Section \ref{Sec:Constraints}, this will have important consequences in one's ability to place constraints on the parameters $\alpha$ and $\gamma$.

Taking the ratio of the power emitted between dark dipole radiation and the gravitational radiation,
\begin{equation}
\frac{P_\text{dark}}{P_\text{GW}} \approx \frac{5}{48} \left(\frac{\gamma}{v^2}\right) \theta(\lambda \omega - 1),
\end{equation}
we see the dipole corrections will be largest early in the inspiral phase, immediately following the activation of the step function.  This will manifest as a negative PN correction to the gravitational waveform.

The inclusion of dipole radiation will not change the orbital frequency or the total energy of the system.  Instead, correcting the balance law to include the dark radiation $-\frac{d}{dt}E_\text{tot} = P_\text{GW} + P_\text{dark}$ will introduce an additional factor to the evolution of the orbital separation in Eq.~(\ref{EQ:DiffRFull}).  Using Eq.~(\ref{EQ:OrbFreq}), the equation for $\dot{r}$ can be rewritten as an equation for the time derivative of the orbital frequency.  Including the dark dipole radiation term, $\dot{\omega}$ can be found in terms of the orbital separation
\begin{widetext}
\begin{equation}
\omega \dot{\omega} = \frac{96\eta m^4}{5r^7}\left[1-\alpha\left(1+\frac{r}{\lambda}+\frac{r^2}{3\lambda^2}\right)e^{-r/\lambda}\right]\left(\frac{\left[1-\alpha \left(1+\frac{r}{\lambda}\right)e^{-r/\lambda}\right]^3}{1-\alpha (1+\frac{r}{\lambda}-\frac{r^2}{\lambda^2})e^{-r/\lambda}}\right)\left(1+ \frac{5\gamma r}{48 m}\frac{\theta(\lambda \omega - 1)}{1-\alpha\left(1+\frac{r}{\lambda}\right)e^{-r/\lambda}}\right).\label{EQ:DarkRadOmdOm}
\end{equation}
\end{widetext}
As we will see in Section \ref{Sec:Waveform}, this equation for the orbital frequency evolution will be necessary when calculating the gravitational waveform.  In particular, the waveform will acquire separate terms for the Yukawa corrections and the dipole radiation corrections, which can be used to constrain the parameters $\alpha$ and $\gamma$ as a function of the Yukawa length scale.

\subsection{Connection to Scalar-Tensor theory}
\label{Sec:BDTheory}

While we have primarily consider the Yukawa potential and dipole radiation in the context of a dark matter model, the kinematic corrections described above are a general feature of most fifth force models.  Scalar-tensor theories have received a lot of attention, in part due to its connection with string theory\cite{Damour:1994ya}. Scalar-tensor theories are a modification to general relativity where an additional scalar degree of freedom is coupled to the trace of the energy momentum tensor (in the Jordan frame), and have been shown \cite{Will:1994fb,Alsing:2011er,Sagunski:2017nzb} to produce the same Yukawa and dipole modification considered here.

In these theories, the ``charge'' accumulation is not due to the accretion of charged particles, but instead a scalar field dependent variation of the inertial mass $m_a(\phi)$ of the compact object \cite{Alsing:2011er}.  When the scalar field theory is written in the Jordan frame \cite{Will:1994fb,Alsing:2011er,Mirshekari:2013vb,Berti:2018cxi}, the dipole radiation and Yukawa corrections can be written in terms of the sensitivity of the body,
\begin{equation}
s_a = -\frac{\partial \log m_a}{\partial \phi}\Bigr\rvert_{\phi_0}.
\end{equation}
In particular, the $\gamma$ parameter for dipole radiation can be written in terms of the sensitivities as
\begin{equation}
\gamma_{ST} = (s_1 - s_2)^2 \left[\frac{2\left(1-\frac{s_1+s_2-2s_1s_2}{2+\omega_{BD}}\right)^2}{2+\omega_{BD}}\right],
\end{equation}
where $\omega_{BD}$ is the Brans-Dicke coupling constant.  One can recover General Relativity by taking $\omega_{BD}\rightarrow \infty$, and thus, using that $\omega_{BD} > 40,000$ from observations of the Shapiro time delay with the Cassini spacecraft~\cite{Bertotti:2003rm}, we can approximate
\begin{equation}
\gamma_{ST} \sim \frac{2  (s_1 - s_2)^2}{\omega_{BD}}\,.
\end{equation}
The additional factor in the square bracket arises from the sensitivity dependence in the gravitational constant, as well as a conversion between scalar ``charge,'' defined in the Einstein frame and the sensitivities, defined in the Jordan frame.

\section{The Gravitational Waveform}
\label{Sec:Waveform}

We now consider the gravitational waveform using the standard amplitude from General Relativity, and apply the results to the case of a binary system with some dark charge.  In principle, corrections to the response function will also arise from additional gravitational wave polarizations that may be sourced by the dark sector we consider in this paper; however, since multiple detectors (or a space-based detector) are needed to detect such additional polarizations, we will neglect them here. We will follow the methods described in \cite{Allen:2005fk,Yunes:2009yz}.  

The plus and cross polarizations of a gravitational wave in General Relativity are given by
\begin{align}
h_+ (t) &= - \left(\frac{1+\cos^2 \iota}{2}\right)\mathcal{A}(t)\cos\left(2\phi_c + 2\phi\left(t-t_c;m,\eta\right)\right),\\
h_\times (t) &= -\left(\cos \iota\right) \mathcal{A}(t)\sin\left(2\phi_c + 2\phi\left(t-t_c;m,\eta\right)\right),
\end{align}
where the gravitational wave amplitude in the time domain is
\begin{equation}
\mathcal{A}(t) = \frac{4\eta m}{D_L}\omega^2(t) r^2(t)\,,
\end{equation}
and where the prefactor is a geometric factor related to the inclination angle $\iota$, i.e.~the angle between the angular momentum of the binary and the observer, while $t_c$ and $\phi_c$ are the time and phase of the binary at coalescence, with $\phi$ the orbital phase of the binary at some time, found by integrating the orbital frequency.

A given detector will have different response functions $F_{+}$ and $F_{\times}$ to the different plus- and cross-polarizations of gravitational waves, which will depend on some additional geometric factors.  In the case of second-generation ground-based instruments, the timescale on which these functions change is much larger than the gravitational wave signal, and thus, they can be treated as constant.  The strain induced on the detector is then given by
\begin{align}
h(t) &= F_+ h_+ (t+t_c-t_0) + F_\times h_\times (t+t_c-t_0),\\
&= -\mathcal{A}(t+t_c-t_0)\bigg[\left(\frac{1+\cos^2\iota}{2}\right)F_+ \cos 2\bar{\phi}(t)\nonumber\\
&\;\;\;\; +\cos\iota F_\times \sin 2\bar{\phi}(t)\bigg],
\end{align}
where $t_0$ is the time when the detector records the coalescence, and $\bar{\phi}(t) \equiv \phi_c + \phi (t-t_0)$.  The strain can be rewritten as a single oscillating function by incorporating the geometric functions into a shift in the phase and a deviation in the luminosity distance:
\begin{align}
D_{\text{eff}} &= D_L \left[F_+^2\left(\frac{1+\cos^2\iota}{2}\right)^2 + F_\times^2 \cos^2\iota\right]^{-1/2},\\
\phi_0 &= \phi_c - \arctan\left(\frac{2\cos\iota}{1+\cos^2\iota}\frac{F_\times}{F_+}\right).
\end{align}
The strain is then given as the function
\begin{equation}
h(t) = -\frac{4\eta m}{D_\text{eff}}\omega^2 r^2\cos\left(2\phi_0 + 2\phi\left(t-t_0;m,\eta\right)\right).
\end{equation}

A matched filtering calculation requires that we compute the Fourier transform of the time-domain waveforms, which can be estimated in the stationary phase approximation.  The Fourier transform of the strain can be written as
\begin{align}
\tilde{h}(f) = -\frac{2\eta m}{D_\text{eff}}\int^{\infty}_{-\infty}&{\dd t}\omega^2 r^2 \Big(e^{i(2\phi_0+2\phi(t) - 2\pi f t)}\nonumber\\
&+e^{-i(2\phi_0+2\phi(t) + 2\pi f t)}\Big),
\end{align}
where the cosine has been expanded in exponentials.  We note that the orbital frequency is monotonically increasing and a positive function (for all cases we consider), properties inherited by $\phi(t)$.

The stationary point is defined as the time $t_s$ when $\omega(t_s) \equiv \dot{\phi}(t_s) = \pi f$. The stationary phase approximation allows one to compute the integral as
\begin{align}
\tilde{h}(f) &= -\frac{2\eta m}{D_\text{eff}}(\pi f)^2r^2(t_s)\left(\frac{\pi}{|\dot{\omega}(t_s)|}\right)^{1/2}\label{EQ:FourH}\\
&\times \exp\left[-i\left(2\pi f t_s - 2\phi_0 - 2\phi(t_s)-\frac{\pi}{4}\sgn(\dot{\omega}(t_s))\right)\right],\nonumber 
\end{align}
where we expect $\sgn(\dot{\omega}(t_s))=1$ in all cases we consider.  One is then required to find the functions $r(t_s), \phi(t_s)$, and $t_s$ as a function of the Fourier frequency.  To find the remaining functions in the phase, we define the quantity $\tau(\omega) = \omega/\dot{\omega}$.  The functions $\phi$ and $t$ can then be rewritten as
\begin{equation}
\phi(\omega) = \int^\omega \tau(\omega') \dd{\omega'} , \;\;\;\;\;\; t(\omega) = \int^\omega \frac{\tau(\omega')}{\omega'} \dd{\omega'}. \label{EQ:FuncDef}
\end{equation}
The binary's phase and time can then be found by $\phi(\omega(t_s))= \phi(\pi f)$ and $t_s=  t(\pi f)$.  Therefore, once the functions $r(\omega)$ and $\dot{\omega}$ are computed for a given model, Eq.~(\ref{EQ:FourH}) can be applied to find the gravitational waveform.

\subsection{Small deformation}

Although the function $\dot{\omega}(r)$ is given in Eq.~(\ref{EQ:DarkRadOmdOm}), the calculation of the orbital separation $r(\omega)$ requires the inversion of Eq.~(\ref{EQ:OrbFreq}).  The relative strength of the Yukawa potential $\alpha$ must be smaller than unity in order for the binary to merge.  Furthermore, to remain consistent with the linear (in $\alpha$) expansion of the potential in Eq.~(\ref{EQ:TotPot}), we wish to find a solution for $r(\omega)$ to linear order in $\alpha$.  Such a solution will correspond to a small General Relativity deformation limit.  This inversion can be done to find the separation, and subsequently $\dot{\omega}(\omega)$, as
\begin{widetext}
\begin{gather}
r(\omega) = \left(\frac{m}{\omega^2}\right)^{1/3}\left[1-\frac{\alpha}{3}\left(1+\frac{m}{\lambda}(m\omega)^{-2/3}\right)\exp\left(-\frac{m}{\lambda} (m\omega)^{-2/3}\right)+\mathcal{O}(\alpha^2)\right], \label{EQ:SCRadSol}\\
\dot{\omega} = \frac{96}{5}{\cal{M}}^{5/3}\omega^{11/3}\left[1-\frac{2\alpha}{3}\left(1+\frac{m}{\lambda}(m\omega)^{-2/3}+\frac{2m^2}{\lambda^2}(m\omega)^{-4/3}\right)\exp\left(-\frac{m}{\lambda}(m\omega)^{-2/3}\right) + \frac{5\gamma}{48}(m\omega)^{-2/3}\theta(\lambda\omega - 1)\right],\label{EQ:DOmDarkRad}
\end{gather}
\end{widetext}
where ${\cal{M}} = \eta^{3//5}m$ is the chirp mass, and the time derivative of the orbital frequency is found by expanding Eq.~(\ref{EQ:DarkRadOmdOm}) to linear order in $\alpha$ where the orbital  separation is evaluated with Eq.~(\ref{EQ:SCRadSol}).

In the inversion of Eq.~(\ref{EQ:DOmDarkRad}), we have dropped terms of  $\mathcal{O}(\alpha\gamma)$.  Neutron stars should naturally accumulate relatively small charge-to-mass rations $\ratqm \ll 1$, hence $\gamma \leq \ratqm^2 \ll 1$.  Explicitly, in order to expand the amplitude and phase of the waveform in Eq.~(\ref{EQ:FourH}) to linear order in $\gamma$, we will require
\begin{equation}
\gamma \ll 12.5 \left(\frac{m}{M_\odot}\right)^{2/3}\left(\frac{\lambda}{1\text{ km}}\right)^{-2/3},\label{EQ:GammaCond}
\end{equation}
so that the dipole radiation term is again a small correction to the General Relativity limit.

Under these conditions, Eq.~(\ref{EQ:FourH}) can be applied to give the Fourier space waveform:
\begin{widetext}
\begin{align}
\tilde{h}(f) &= -\left(\frac{5\pi}{24}\right)^{\frac{1}{2}}\frac{{\cal{M}}^2}{D_\text{eff}}(\pi {\cal{M}} f)^{-\frac{7}{6}}\left[1-\frac{\alpha}{3}\left(1+\frac{m}{\lambda}(\pi m f)^{-\frac{2}{3}} - \frac{2 m^2}{\lambda^2} (\pi m f)^{-\frac{4}{3}}\right)\exp\left(-\frac{m}{\lambda}(\pi m f)^{-\frac{2}{3}}\right)\right.\nonumber\\
&\;\;\;\; \left. - \frac{5\gamma}{96}(\pi m f)^{-\frac{2}{3}}\theta(\pi \lambda f - 1)\right]e^{-i\Psi},\label{EQ:SmallCoupFourH}\\
\Psi &= 2\pi f t_0 - 2\phi_0 - \frac{\pi}{4} + \frac{3}{128}\left(\pi {\cal{M}} f\right)^{-5/3}\left[1+\frac{20\alpha}{3}F_3\left(\frac{m}{\lambda} (\pi m f)^{-2/3}\right) - \frac{5\gamma}{84}(\pi m f)^{-\frac{2}{3}}\theta(\pi\lambda f - 1)\right],\label{EQ:SmallCoupPh}
\end{align}
\end{widetext}
where we have defined
\begin{align}
F_3(x) &=\left(\frac{180 + 180 x + 69 x^2 + 16 x^3 + 2 x^4}{x^4}\right)e^{-x} 
\nonumber \\ &+ \frac{21\sqrt{\pi}}{2x^{5/2}}\erf({\sqrt{x}}), 
\end{align}
and erf$(x)$ is the error function\footnote{The error function can be represented approximately by 
\begin{equation}
\erf(\sqrt{x})\approx 1 - \left(1+a_1 x^{1/2} + a_2 x + a_3 x^{3/2} + a_4 x^2\right)^{-4}\,,\nonumber
\end{equation}
with $a_1 = 0.278393, a_2 = 0.230389, a_3 = 0.000972, a_4 = 0.078108$, if one wishes.}.

We see that inclusion of dipole radiation manifests as a -1PN correction.  The magnitude of this contribution can become very large at early times, however the step-function modulates this behavior by abruptly shutting off the contribution when Eq.~(\ref{EQ:DipActive}) is not satisfied. In contrast, the Yukawa-type modifications to the waveform do not easily separate into a post-Newtonian expansion as a functions of $x=\frac{m}{\lambda}(\pi m f)^{-2/3}$.  Both the amplitude and phase functions remain bounded for all positive (physical) values of $x$, thus these corrections remain well behaved throughout the binary inspiral.

\subsection{Mass range of the dark photon}

If we could observe the inspiral over its entire evolution (starting at infinite separation), $\frac{m}{\lambda}(m\omega)^{-2/3}$ would start arbitrarily large and eventually decay to the limit where $r\ll \lambda$.  In this scenario, one needs to use the full waveform found in Eq.~(\ref{EQ:SmallCoupFourH}) and Eq.~(\ref{EQ:SmallCoupPh}) in order to properly incorporate the non-perturbative behavior of the solutions.  During the inspiral phase, however, the binary will emit gravitational waves at low frequencies for a longer period of time than at higher frequencies.  For observations beginning at a gravitational wave frequency $f_0$, we ca then look at the limiting behavior of the waveform when $x_{0} \gg 1$ (the heavy limit) and when $x_{0} \ll 1$ (the ultra-light limit), where we have defined $x_0 \equiv \frac{m}{\lambda}(\pi m f_0)^{-2/3}$.  In these limiting studies, we ignore the dipole radiation term, as it remains uncoupled from the Yukawa corrections, and does not simplify in any limit involving $x_0$.

As we will see, degeneracies arise in the limiting regimes which are not present in the full waveform.  These degeneracies will play an important role in our ability to constrain the relative Yukawa strength $\alpha$ in Section \ref{Sec:Constraints}.

\subsubsection{A heavy dark photon}

For sufficiently large dark photon masses, $x_0 \gg 1$ throughout the observational window.  In this case, the nonperturbative exponential functions suppress these corrections below any detectable range, as these terms remain proportional to $e^{-x_0}$.  In this regime, the amplitude of the waveform, given by Eq.~(\ref{EQ:SmallCoupFourH}), does not acquire any corrections to linear order in $\alpha$.  The phase in Eq.~(\ref{EQ:SmallCoupPh}) only receives linear $\alpha$ corrections from the error function.  However, one can see from the integral definition,
\begin{equation}
\erf\left(\sqrt{x_0}\right) = \frac{2}{\sqrt{\pi}}\int_0^{\sqrt{x_0}} e^{-t^2}dt \;\; \rightarrow \;\; 1 + \frac{2}{\sqrt{\pi}}e^{-x_0}+... 
\end{equation}
that the only non-exponential correction from the error function will be a constant, degenerate with the phase $\phi_0$.  As a result, the Yukawa corrections for a heavy dark photon becomes completely degenerate with the General Relativity waveform.

\subsubsection{An ultra-light dark photon}
\label{Sec:ULPhoton}

We now consider the case where observation of the binary begins after the binary has entered the range of the Yukawa interaction.  In this case, $r\ll \lambda$, and the Yukawa potential can be Taylor expanded.  Of course, this implies that we cannot take the infinite orbital separation limit and that the above condition will only be satisfied for a set of masses.  This condition can be explicitly written in terms of the Yukawa length scale $\lambda$, or equivalently the dark photon mass, as
\allowdisplaybreaks[4]
\begin{align}
\lambda &\gg \left(520\text{ km}\right)\left(1-\frac{\alpha}{6}\right)\left(\frac{f_0}{10 \text{ Hz}}\right)^{-\frac{2}{3}}\left(\frac{m}{M_\odot}\right)^{\frac{1}{3}}, \label{EQ:SmallMv}\\
m_v &\ll \left(3.8 \times 10^{-13}\text{ eV}\right)\left(1+\frac{\alpha}{6}\right)\left(\frac{f_0}{10 \text{ Hz}}\right)^{\frac{2}{3}}\left(\frac{m}{M_\odot}\right)^{-\frac{1}{3}}.
\end{align}
Due to the extremely light mass required for the dark photon, we call this the \emph{ultra-light dark photon} limit, corresponding to $x_0\ll 1$.  In this limit, the gravitational waveform can be written as
\begin{widetext}
\begin{align}
\tilde{h}_\text{ul}(f) &= -\left(\frac{5\pi}{24}\right)^{\frac{1}{2}}\frac{{\cal{M}}^2}{D_\text{eff}}\left(\pi {\cal{M}} f\right)^{-\frac{7}{6}}\left[1-\frac{\alpha}{3} +\frac{5\alpha m^2}{6\lambda^2}(\pi m f)^{-\frac{4}{3}}-\frac{7\alpha m^3}{9\lambda^3}(\pi m f)^{-2} + \mathcal{O}\left(\frac{m^4}{\lambda^4}(\pi mf)^{-\frac{8}{3}}\right)\right] e^{-i\Psi_\text{ul}},\label{EQ:AmpCheckSDPM}\\
\Psi_\text{ul} &= 2\pi f t_0 - 2\phi_0 - \frac{\pi}{4} + \frac{3}{128}\left(\pi {\cal{M}} f\right)^{-5/3}\left[1+\frac{2\alpha}{3}+\frac{10\alpha m^2}{27\lambda^2}(\pi m f)^{-\frac{4}{3}}-\frac{200\alpha m^3}{693\lambda^3}(\pi m f)^{-2} + \mathcal{O}\left(\frac{m^4}{\lambda^4}(\pi mf)^{-\frac{8}{3}}\right)\right].\label{EQ:PhaseCheckSDPM}
\end{align}
\end{widetext}

The Fourier amplitude does not pick up any corrections to first order in the dark photon mass.  The two paramount functions for calculating the amplitude and phase, Eq.~(\ref{EQ:OrbFreq}) and Eq.~(\ref{EQ:DarkRadOmdOm}), only contain corrections of the form $\left(1+\frac{r}{\lambda}+\mathcal{O}\left(\frac{r}{\lambda}\right)^2\right)\exp\left(-\frac{r}{\lambda}\right)$.  Taking the $r\ll \lambda$ expansion of these equations will result in no linear order correction.  This property is inherited by the separation function during the inversion of Eq.~(\ref{EQ:OrbFreq}) due to the term-by-term matching of the perturbative series. 

The leading order correction in both the phase and amplitude appears at -2PN, with corrections to this appearing at \emph{more negative post-Newtonian orders}. This is consistent again with the expansion requirements of this section, namely $r \ll \lambda$. One can for example check that the -2PN order term is actually larger than the -3PN order term because $\frac{m}{\lambda} \ll (\pi m f)^{2/3} \sim v^{2} \sim m/r$. Therefore, when including $\lambda$ corrections, the usual post-Newtonian order counting is not applicable. Instead, the model presented above is a \emph{bivariate expansion} in both $v \ll 1$ and $r \ll \lambda$.

We note that the first correction to both the amplitude and phase of the waveform is independent of $\lambda$.  This introduces a degeneracy between the chirp mass and the Yukawa strength parameter $\alpha$.  It is ultimately this degeneracy that is explored in \cite{Croon:2017zcu}.  This degeneracy is lifted by the -2PN correction.  However, both amplitude and phase depend only on the quantity $\alpha m^{2/3}\lambda^{-2}$, which implies there is a 100\% degeneracy between $\alpha$ and $\lambda$. This degeneracy is lifted when we include the -3PN correction, which depends on a different combination of $\alpha$ and $\lambda$. This is analogous to the degeneracy between the component mass $m_{1}$ and $m_{2}$ in General Relativity at Newtonian order, which is lifted when one includes 1PN corrections.

\subsection{Relative magnitude of Yukawa and dipole corrections}

We now consider the region of parameter space where the dipole radiation modifications of the waveform dominate over the Yukawa modifications.  Due to the particular sensitivity of gravitational wave interferometers to the phase of the gravitational wave, we focus on the phase modifications presented in Eq.~(\ref{EQ:SmallCoupPh}).  The dipole radiation modifications will be dominant under the condition
\begin{equation}
\frac{5\gamma}{84\alpha v^2}\theta\left(\frac{\lambda}{m}v^3-1\right) \geq \frac{20}{3}F_3\left(\frac{m}{\lambda v^2}\right).
\end{equation}
The requirement of a valid post-Newtonian expansion ($v\ll 1$) can be combined with the requirement that the step-function condition is satisfied to find
\begin{equation}
\frac{m}{\lambda v^2} \leq v \ll 1,
\end{equation}
which corresponds to the ultra-light dark photon limit.  Therefore, after removing boundary terms, the dipole corrections to the waveform are only present in the waveform from Eq.~(\ref{EQ:AmpCheckSDPM}) and Eq.~(\ref{EQ:PhaseCheckSDPM}).  The condition that dipole radiation dominates over the Yukawa modifications can be rewritten as
\begin{equation}
v^2 \leq \frac{5\gamma}{56\alpha}+\mathcal{O}\left(\frac{m^2}{\lambda^2 v^4}\right), \;\;\;\; \text{ and} \;\;\;\; v^3 \geq \frac{m}{\lambda}. \label{EQ:GModDomCond}
\end{equation}
The second of these conditions is precisely the condition in Eq.~(\ref{EQ:DipActive}), requiring the step-function to be active.  The only significant deviations from these approximate requirements come when the orbital velocity approaches unity, which also allows the minimum $\lambda/m$ to approach unity. In this regime, of course, the post-Newtonian expansion is valid no longer and a full numerical analysis is required.

\section{Constraints on Dark Matter Model Parameters}
\label{Sec:MethodofConstraints}

\subsection{Fisher analysis basics}
\label{basics}

The Fisher information matrix is a standard statistical tool used to estimate the accuracy to which parameters can be measured in gravitational wave physics in the large signal-to-noise ratio limit (for a detailed discussion, see e.g.~\cite{Porter:2015eha, Vallisneri:2007ev}). The inverse of the Fisher information provides a lower bound on the error of any unbiased estimator (the Cramer-Rao bound), and hence provides an optimistic set of forecasted constraints, as compared to a Bayesian analysis. The appeal of this approach is the computational efficiency; it requires orders of magnitude less computing power then a Markov-Chain Monte Carlo analysis.

The Fisher information matrix $\Gamma_{ab}$ is defined as a weighted inner product of derivatives of the waveform with respect to parameters $\theta^a$ and $\theta^b$. That is,
\be
\label{FIM}
\Gamma_{ab} \equiv \left( \frac{\partial h}{\partial \theta^a} \Bigg| \frac{ \partial h }{\partial \theta^b} \right) ,
\ee
where the inner product is defined as
\be
\label{innerproduct}
\left( h_1 | h_2 \right) \equiv 2  \int _{f_{\rm low}} ^{f_{\rm high}} \frac{\tilde{h}_1 \tilde{h}_2^* + \tilde{h}_1^* \tilde{h}_2}{S_n(f')} {\rm d}f' ,
\ee
with $S_n(f)$ the spectral noise density of the detector, and $\tilde{h}(f)$ the Fourier transform of the time-domain response $h(t)$. From this definition, one can quickly see that the signal to noise ratio (SNR) is given by
\be
\rho^2 \equiv \left( h | h\right) = 4\int {\rm dlog}f\; \frac{f|\tilde{h}|^2}{S_n(f)} . \label{EQ:defSNR}
\ee
The bounds of integration in Eq.~\eqref{innerproduct} are discussed in detail in Section \ref{range}.

The Fisher matrix is equivalent to evaluating the second derivative of the likelihood ${\cal{L}}$
\be
\Gamma_{ab} = - {\rm E} \left[ \frac{\partial^2 {\cal L}}{\partial \theta^a \partial \theta^b}\right] ,
\ee
at the maximum likelihood estimate for $\theta^a$, where ${\cal{L}}$ is given by
\be
{\cal L}({\bf \theta})  = \exp \left[- \frac{1}{2} \left( s - h({\bf \theta}) | s - h({\bf \theta}) \right)\right],
\ee
given a signal $s$ and a gravitational waveform $h$.  Hence, the inverse of the Fisher matrix can alternatively be viewed as the frequentist error of the maximum likelihood estimator. A third interpretation of the Fisher information matrix is a Bayesian one: the inverse Fisher matrix is the covariance of the posterior probability distribution of the true parameters, as would be inferred by a Bayesian analysis of a single experiment, assuming constant prior probabilities, a high SNR, and Gaussian noise.

From these definitions, one can estimate the sensitivity of a detector to a given parameter. The root-mean-squared (i.e.~1$\sigma$) error on a parameter $\theta^a$ can be estimated by,
\be
\Delta \theta^a \leq \sqrt{\Sigma^{aa}} ,
\ee
where $\Sigma^{aa}$ is defined as the $(a,a)$ component of the covariance matrix $\Sigma^{ij}\equiv (\Gamma_{ij})^{-1}$. In this work we will use Eqs.~\eqref{FIM} and \eqref{innerproduct} to compute the above error, which we interpret as the projected sensitivity of a given detector to a parameter $\theta^a$.  To prevent (numerically) singular Fisher matrices, we follow the method of \cite{Chamberlain:2017fjl}, where we use a working precision of one hundred decimal places and invert the Fisher matrix by the Cholesky decomposition.

\subsection{Range of frequency integration}
\label{range}

The limits of integration in the Fisher analysis dictate the range over which our waveform in Eq.~(\ref{EQ:SmallCoupFourH}) remains valid and detectable above detector noise.  For the detectors we consider (see Section \ref{Sec:Detectors}), typical binary neutron star and mixed black hole-neutron star inspirals will merge within the detector's frequency window.  The high frequency limit will then remain independent of the particular detector, given instead by physical quantities of the binary.  However, the low frequency limit will depend on the sensitivity of a particular detector.

For the low frequency limit, we follow \cite{Chamberlain:2017fjl}, defining
\begin{equation}
f_{\rm low} = \max \left[f_{\rm low-cut}, f_{\rm lratio}\right],
\end{equation}
where $f_\text{low-cut}$ is a detector dependent cutoff frequency given as $1$ Hz for the Einstein Telescope (ET), and $5$ Hz for the remaining detectors we consider in Section \ref{Sec:Detectors}.  The frequency $f_\text{lratio}$ is defined as the lowest frequency where the amplitude of the gravitational wave signal is $10\%$ of the detector noise spectrum.  Below this frequency, the integrand in Eq.~(\ref{EQ:defSNR}) is less than $\mathcal{O}(10^{-2})$, and can thus we neglected when computing the signal-to-noise ratio.

At high frequencies, our waveform becomes invalid \cite{Ajith:2007kx,Abbott:2016bqf} due to a lack of stable circular orbits, assumed in the orbital frequency in Eq.~(\ref{EQ:OrbFreq}) and the complete breakdown of the post-Newtonian approximation.  The frequency of gravitational waves \cite{Giudice:2016zpa} emitted at the innermost stable circular orbit (ISCO) (for a test particle in a Schwarzschild spacetime of mass $m$) is given by
\begin{equation}
f_\text{ISCO} = (4.4\times 10^3 \text{ Hz})\left(\frac{m}{M_\odot}\right)^{-1}.
\end{equation}
However, when the binary contains a neutron star, the waveform must be terminated before contact.  The contact frequency \cite{Lehner:2016lxy} can be approximated as the gravitational wave frequency at which the separation is equal to the sum of the radii of the two stars:
\begin{align}
f_\text{contact} &= (4.4\times 10^3 \text{ Hz})\left(\frac{m}{M_\odot}\right)^{-1}(6\tilde{C})^{3/2} \\ 
\tilde{C}^{-1} &= \frac{m_1}{m C_1} + \frac{m_2}{m C_2},
\end{align}
where $C_i$ is the compactness of the $i$th star\footnote{For a non-rotating black holes, the compactness is taken as $C = \frac{1}{2}$, while neutron stars have the upper bound  $C\leq \frac{4}{9}$. For stable neutron stars, the compactness \cite{Yagi:2013awa} is typically in the range $C \in (0.1,0.2)$.}, and $\tilde{C}$ acts as an effective compactness for the binary.  The high frequency limit must be taken as the minimum between these two frequencies,
\begin{equation}
f_\text{high} = \min\left[f_\text{contact},f_\text{ISCO}\right].
\end{equation}

As discussed in \cite{Mandel:2014tca}, enforcing this high frequency cut-off can lead to incorrect results for the accuracy of parameter-estimation. This particularly affects parameters that depend sensitively on the merger time, such as the total mass, and thus is particularly relevant for higher mass systems. In contrast, the accuracy to which dipole and Yukawa modifications can be constrained builds up during the early inspiral phase, and further, in this work we study only low mass systems. Hence we do not expect parameter-estimation to depend sensitively on the merger phase.

\begin{figure}
\hspace{-3mm}
\begin{center}
\includegraphics[width=0.48 \textwidth]{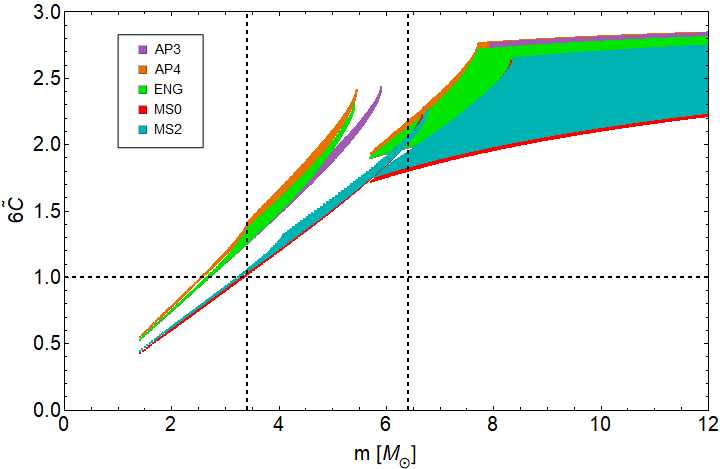}
\caption{Effective compactness of neutron star binaries and mixed black hole-neutron star binaries, for various neutron star equations of state \cite{Yagi:2013awa}. We have taken the minimum black hole mass as $5 M_\odot$.  The vertical, dashed black lines correspond to the total masses we will consider in Section \ref{Sec:Constraints}.  In both cases, we see that $6\tilde{C} \geq 1$.}\label{Fig:Comp}
\end{center}
\vspace{0cm}
\end{figure}

We can further simply $f_{\rm high}$ as follows. Stable neutron stars have roughly the same radius $R_\text{NS}$, given by their equation of state \cite{Yagi:2013awa}, and therefore the compactness of the individual star is given by $C_i \approx m_i/R_\text{NS}$.  The effective compactness $\tilde{C}_\text{NS-NS}$ can then be rewritten as
\begin{equation}
6\tilde{C}_\text{NS-NS} \approx \frac{3m}{R_\text{NS}} = 0.44 \left(\frac{m}{M_\odot}\right)\left(\frac{R_\text{NS}}{10 \text{ km}}\right)^{-1}.\label{EQ:CNSNS}
\end{equation}
Similarly, we take the black hole compactness to be $C = \frac{1}{2}$, so the effective compactness for a black hole - neutron star binary can be written as
\begin{equation}
6\tilde{C}_\text{BH-NS} \approx \frac{6 m}{2m_\text{BH} + R_\text{NS}}.
\end{equation}
For a particular neutron star equation of state, the effective compactness for various binaries systems can be calculated.  Figure~\ref{Fig:Comp} displays the range of $6\tilde{C}$ for various neutron star equations of state.  We see that the effective compactness is greater than unity for all but low mass binary neutron stars.

When $6\tilde{C} > 1$, the contact frequency occurs after $f_\text{ISCO}$.  Therefore, the high frequency limit is written as
\begin{align}
f_\text{high} &= (4.4\times 10^3 \text{ Hz})\left(\frac{m}{M_\odot}\right)^{-1}(6\mathcal{C})^{3/2},\label{EQ:fhDef}\\
\mathcal{C} &= \min\left[\tilde{C},\frac{1}{6}\right].
\end{align}
For the binaries we consider in Section \ref{Sec:Constraints}, $6\mathcal{C} = 1$, thus our analysis will always take the high frequency limit as $f_\text{ISCO}$.

The mass-radius relations of neutron stars used in Fig.~\ref{Fig:Comp} does not include the effects of a dark matter core.  However, recent work \cite{Ellis:2018bkr} has shown that for a particular equation of state, the same total mass neutron star will typically have a smaller radius when a dark core is included.  This implies that including dark matter will increase the compactness of a particular neutron star, further increasing the effective compactness $6\tilde{C}$.  Therefore, $f_\text{high} = f_\text{ISCO}$ remains valid for the binary systems of interest.

\begin{table*}[]
\begin{tabularx}{1\textwidth}{Y|Y|Y|Y|Y|Y}
   \hline\hline
Name& $m_{1} [M_{\odot}]$ & $m_{2} [M_{\odot}]$ & $(\chi_{1},\chi_{2})$ & $D_\text{eff}$  & SNR (aLIGO)\\ \hline
NSNS&  2.0 	& 	1.4 	& 	(0.01, 0.02) 	& 	 100 Mpc	& 25\\
BHNS & 5.0 	& 	1.4 	&	(0.2 , 0.02)		& 	150 Mpc	& 25\\
   \hline\hline
\end{tabularx}
\caption{Representative systems used in our Fisher analysis.  The signal-to-noise ratio is given for Adv. LIGO at design sensitivity.}
\label{tab:table}
\end{table*}

\subsection{Future detectors and sensitivity curves}
\label{Sec:Detectors}

\begin{figure}
\hspace{-3mm}
\begin{center}
\includegraphics[width=.49 \textwidth]{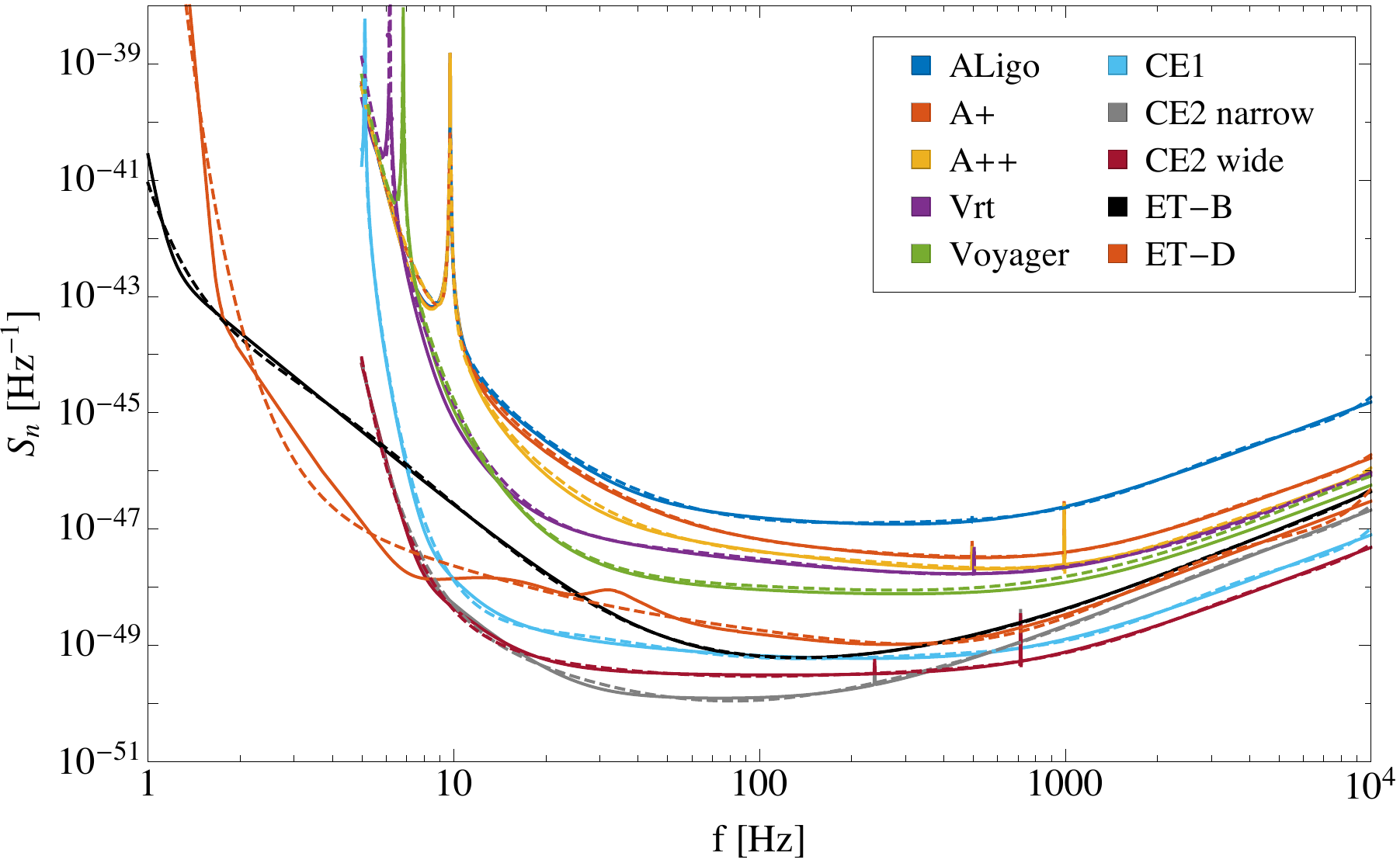}
\caption{Projected spectral noise density (solid) and analytic fits (dashed) for each detector we consider. The curves are truncated at the particular detector's cutoff frequency $f_\text{low-cut}$.}
\label{fig:Fits}
\end{center}
\vspace{0cm}
\end{figure}

In this work we compute forecasted constraints on dark sector modifications for a set of 10 ground-based detectors: aLIGO at design sensitivity \cite{Aasi:2013wya}, aLIGO with squeezing (A+/A++  \cite{LIGOwhitepaper}), Voyager \cite{LIGOwhitepaper}, VRT \cite{Adhikari:2013kya,LIGOwhitepaper}, Cosmic Explorer 1 (CE1) and 2 narrow-band and wide-band configurations (CE2n and CE2w respectively) \cite{LIGOwhitepaper}, and the Einstein Telescope in its single interferometer configuration (ET-B) and in ``xylophone'' configuration ET-D \cite{Hild:2009ns, ET}. 

For a detailed overview of the detector sensitivities we refer the reader to \cite{Evans:2016mbw}. Here we briefly summarize the salient details of each detector: \newline
\newline \noindent {\it A+, A++:} Upgrades to LIGO to minimize quantum and thermal noise, operational starting around 2020.   \newline

\noindent {\it Voyager:} An upgrade to LIGO, which replaces glass mirrors and suspensions with silicon parts, and will operate at a cryogenic temperature of 123K. To be operational in 2027.  \newline
  
\noindent {\it Vrt:} The same as Voyager, but operated at room temperature, instead of at cryogenic temperatures. \newline   
   
\noindent {\it Cosmic Explorer:} Aims to observe binaries at high redshift ($z>1$), using 40km long detectors. CE1 is built on A+ technology, while CE2 (in narrow band and wide band configurations) is built on Voyager technology.  Projected start date of $2035$.\newline
   
\noindent {\it Einstein Telescope:} Designed to improve upon low-frequency ($f < 10$ Hz) noise levels. To be built underground, operational in 2030-2035.  \newline

For each of these detectors, we find an analytic fit to the tabulated projected sensitivities. These fitting functions will greatly accelerate the computation of Fisher matrix elements. The tabulated and analytic fit sensitivity curves are shown in Figure \ref{fig:Fits}. Details of the fits can be found in Appendix \ref{appFits}.

\subsection{Constraints on dark sectors}
\label{Sec:Constraints}

We now apply the Fisher analysis discussed in Sec.~\ref{basics} to the most general waveform, calculated in Eq.~(\ref{EQ:SmallCoupFourH}), and include the General Relativity corrections, up to 2PN order, calculated in \cite{Khan:2015jqa}.  In particular, we will look at a binary neutron star and a mixed black hole-neutron star binary, evaluated at the parameters found in Table \ref{tab:table}.  The maximal list of parameters we consider is given by
\begin{equation}
{\boldsymbol \theta} = \left\{ \log{\cal A}, t_c, \phi_c, \log{\cal M}_c , \log\eta , \chi_s, \chi_a , \alpha, \gamma \right\},\label{EQ:Params}
\end{equation}
where $\chi_s = (\chi_1 + \chi_2)/2$, $\chi_a = (\chi_1 - \chi_2)/2$, and $\chi_i$ is the dimensionless spin parameter for the $i$th star. Our Fisher analysis, thus, will include all covariances between the parameters listed above.  We also note that our set of parameters does not include spin precession or tidal parameters, as these enter at higher PN order.

When projecting future constraints, we will assume that future gravitational wave observations are consistent with General Relativity.     This implies that when computing the Fisher matrix elements, we will take the General Relativity limit $\Gamma_{ab}\rvert_{\alpha,\gamma \rightarrow 0}$. A by-product of this is that we lose the ability to constrain the length scale of the Yukawa interaction $\lambda$ separately, and thus, this parameter does not appear in Eq.~(\ref{EQ:Params}).  This can be seen directly in Eq.~(\ref{EQ:SmallCoupFourH}), noticing that any derivative with respect to $\lambda$ is proportional to either $\alpha$ or $\gamma$.  Instead, the constraints placed on each of these parameters will have a functional dependence on the Yukawa length scale. This has the added benefit that the Fisher analysis will not have numerical errors due to the sharp features manifesting from derivatives of the Heaviside function in dipole radiation.

In the case of the mixed binary, the black hole should not be charged under the massive dark photon \cite{Bekenstein:1971hc}, and thus we expect $\alpha = 0$.  For this reason, we do not include $\alpha$ in the list of parameters when considering the mixed binary in a Fisher analysis.  This parameter, however, could be included in the future as a test of black hole no-hair theorems.  While nonzero $\alpha$ can also be attributed to a dark matter cloud surrounding the black hole, tidal effects may become relevant before the $f_\text{high}$ considered here.

Similarly, when dipole emission is not present in the waveform, the parameter $\gamma$ will be removed from the parameter list. This occurs when Eq.~(\ref{EQ:DipActive}) is not satisfied, forcing the step function to vanish and removing the dipole radiation terms from the waveform.  The parameter $\gamma$ can only be constrained when the step-function is active sometime before the end of the observation, given by the frequency $f_\text{high}$.  Using the definition of $f_\text{high}$ in Eq.~(\ref{EQ:fhDef}), we find the minimum length scale as
\begin{equation}
\lambda \geq \left(22\text{ km}\right)\left(6\mathcal{C}\right)^{-3/2}\left(\frac{m}{M_\odot}\right),
\end{equation}
for which we include $\gamma$ as a parameter in the Fisher analysis.

\begin{figure*}[]
\hspace{-3mm}
\begin{center}
\includegraphics[width=0.96 \textwidth]{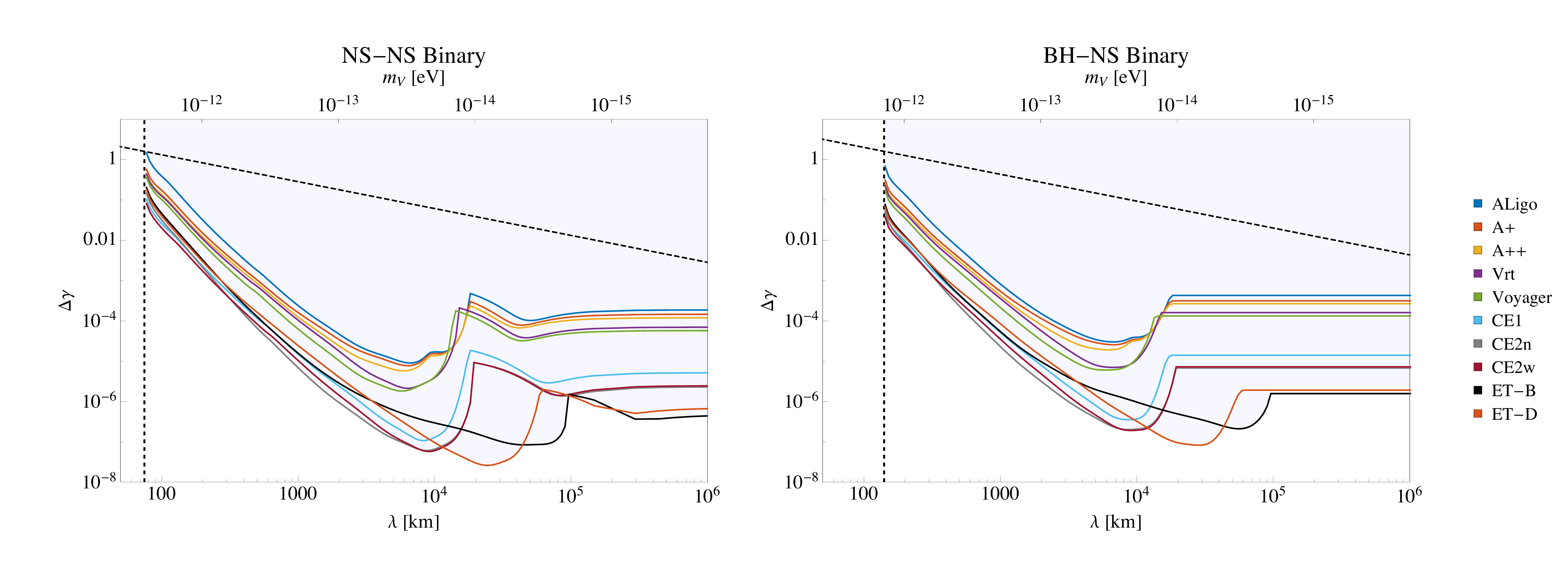}
\vspace{-2mm}
\caption{Projected constraints on the charge asymmetry $\gamma$ with future neutron star binary (NSNS) and mixed black hole-neutron star binary (BHNS) observations.  The vertical dashed line denotes the activation of dipole radiation at some point before $f_\text{ISCO}$.  The diagonal dashed line gives the consistency requirement of the waveform, given by Eq.~(\ref{EQ:GammaCond}).  All detectors considered are able to constrain $\gamma$ from our waveform until $\lambda \sim 10^8$ km.}
\label{fig:gammaplot}
\end{center}
\vspace{0cm}
\end{figure*}

Under these considerations, we estimated projected constraints for $\gamma$ from both binary systems, as shown in Fig.~\ref{fig:gammaplot}.  The earlier dipole radiation activates, the more significant its contribution becomes to the signal-to-noise ratio.    Thus, as the Yukawa length scale increases, the constraint on $\gamma$ becomes more stringent, until $\lambda \sim \mathcal{O}\left(10^4-10^5\text{ km}\right)$.  At this length scale, the step function is activated before the low frequency bound, given by Eq.~(\ref{EQ:DipActive}).  Approximating the low frequency limit as $f_\text{low-cut}$, we find this critical length scale to be $\lambda \approx 10^5$ km for ET, and $\lambda \approx 2\times 10^4$ km for the remaining detectors.  Above this length scale, the length scale $\lambda$ only enters the waveform through the Yukawa-corrections.  For BHNS binaries, we have no Yukawa corrections, thus the constraint is independent of the length scale.  For NSNS binaries, these Yukawa corrections maintain a (weak) lambda dependence, causing the constrain to asymtote to a particular (detector-dependent) value.

\begin{figure*}[]
\hspace{-3mm}
\begin{center}
\includegraphics[scale=.45]{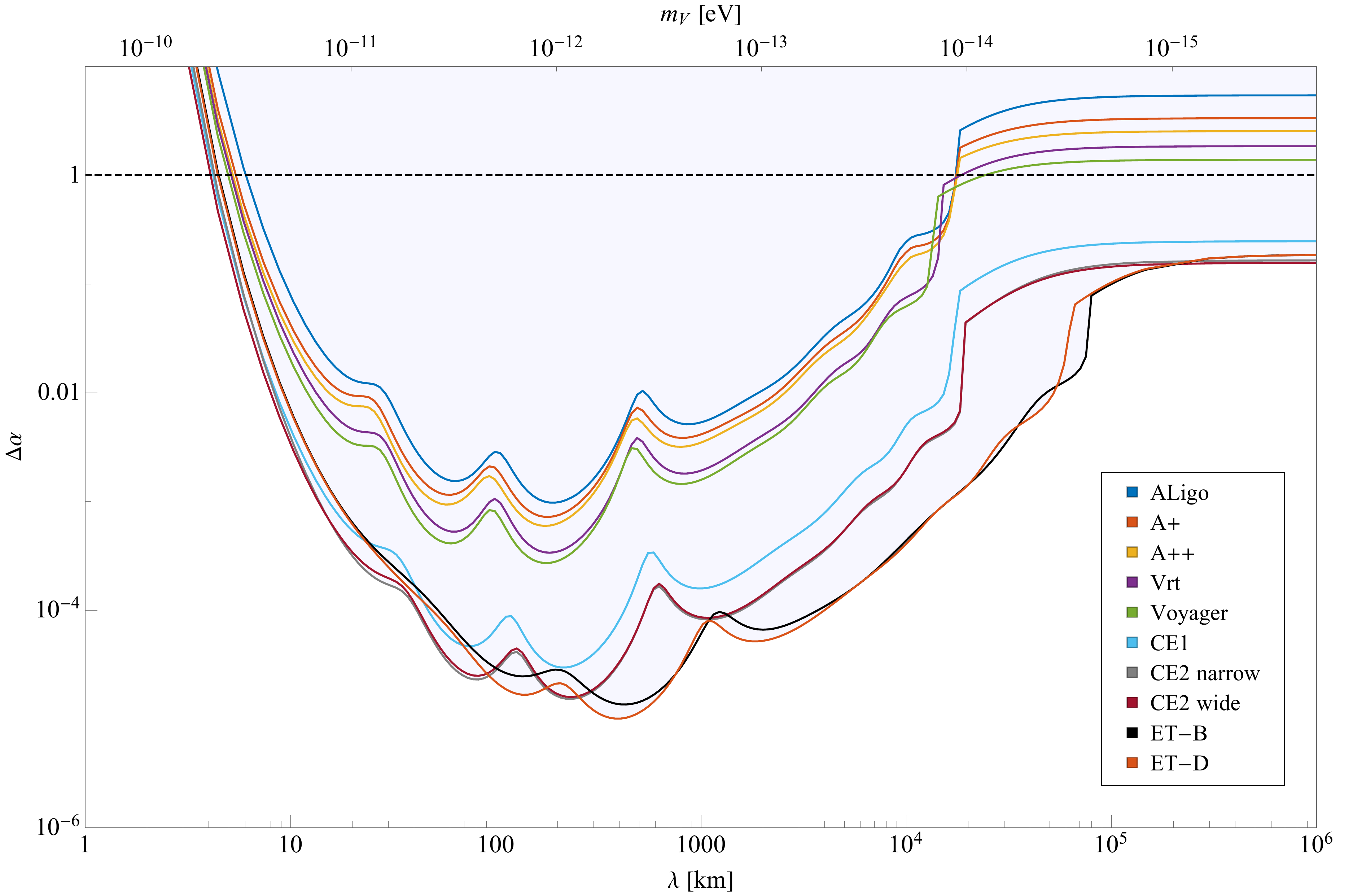}
\caption{Projected constraints on the relative strength of the Yukawa interaction $\alpha$ between neutron stars.  The dashed line at $\alpha = 1$ corresponds to the physical requirement that the total energy in Eq.~(\ref{EQ:Energy}) remains negative throughout the inspiral.}
\label{fig:alphaplotNSNS} 
\end{center}
\vspace{0cm}
\end{figure*}

The relative Yukawa strength $\alpha$ can also be constrained from future binary neutron star observations, as shown in Fig.~\ref{fig:alphaplotNSNS}.  We find that significant constraints can be placed on the relative Yukawa strength above $\lambda \sim 5$ km.  Below this length scale, the exponential suppression of the Yukawa interaction leads to minuscule corrections to the waveform through the inspiral.  Surprisingly, even when the Yukawa length scale is comparable to the radius of the neutron star ($R_{NS}\sim 13$ km), we are still able to constrain $\alpha \leq 10^{-2}$.  Once one crosses into the ultra-light regime, $\lambda \gtrsim \mathcal{O}(10^3\text{ km})$, we again see a rapid decline in the strength of the constraint due to the small Yukawa corrections shown in Eq.~(\ref{EQ:AmpCheckSDPM}).  It is during this regime that the dipole radiation terms can begin to dominate for a significant period of the inspiral phase.  Again, once dipole radiation activates throughout the entire detection window, $\alpha$ is only constrained below the consistency bound $\alpha < 1$ by the more sensitive CE and ET detectors.

\begin{figure}
\hspace{-3mm}
\begin{center}
\includegraphics[width=.49 \textwidth]{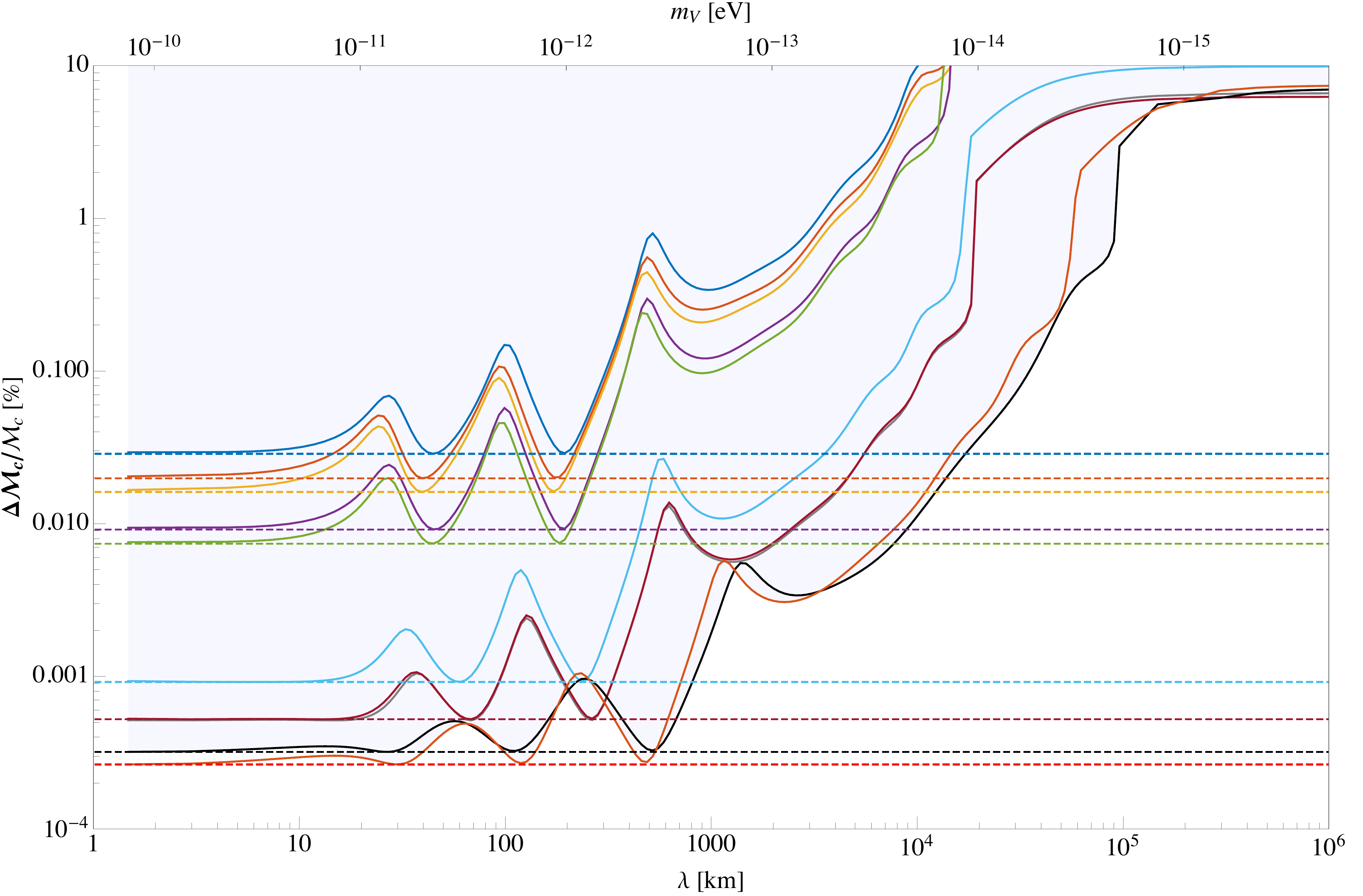}
\caption{Projected sensitivity to the chirp mass in a binary neutron star merger, with and without dark sector modifications. Dashed lines are the sensitivity predicted by the GR waveform, while the solid lines are the sensitivity once dark sectors are included. Colors are as in previous plots.}
\label{fig:Mc}
\end{center}
\vspace{-0cm}
\end{figure}

In the ultra-light regime, $\lambda \gtrsim \mathcal{O}(10^3\text{ km})$, all previous figures show an increase in the variance of estimated parameters, see e.g.~the variance of $\alpha$ in Fig.~\ref{fig:alphaplotNSNS}. The variance of the estimated astrophysical parameters, like the chirp mass, also increases in this regime, as we can see in Fig.~\ref{fig:Mc}, which for illustrative purposes focuses on a NSNS merger. The reason for this increase in the variance is a similar increase in the correlation between the $\alpha$ parameter and the chirp mass; we have indeed verified that this element of the correlation matrix approaches unity as $\lambda \gtrsim \mathcal{O}(10^3\text{ km})$. We can see the growth of this correlation analytically in Eq.~\eqref{EQ:PhaseCheckSDPM}: as $\lambda$ becomes large, the $1/\lambda^{2}$ and the $1/\lambda^{3}$ terms in the Fourier phase become small, and the leading order term in the phase depends not on just the chirp mass, but rather the product of the chirp mass and a $(1 + 2 \alpha/3)$ factor. This makes the Fisher matrix nearly degenerate, which then leads to a very large variance upon inversion. In this regime, parameter estimation with GR templates could be subject to ``fundamental theoretical bias'' \cite{Yunes:2009ke}. 

We now return to dark matter. One can convert the bounds on $\alpha,\gamma$ into an upper bound on the charge to mass ratio $\ratqm$ of the neutron star via
\begin{equation}
\ratqm \leq \frac{\sqrt{\gamma_b}+\sqrt{\gamma_b+4\alpha_b}}{2},\label{EQ:qmConst}
\end{equation}
where $\alpha_b,\gamma_b$ are the bounding functions given in Figs.~\ref{fig:gammaplot} and~\ref{fig:alphaplotNSNS} for a particular detector. This relation follows straightforwardly from the definitions of $\alpha$ and $\gamma$. We note, that for the mixed black hole-neutron star system, the assumption that $\alpha = 0$ provides the stronger constraints $\ratqm\leq\sqrt{\gamma_b}$.  To date, no gravitational wave observations have been made of a mixed binary, so we will focus on the binary neutron star case below instead.

Using the dark matter model described in Sec.~\ref{Sec:Model}, the constraint on the charge-to-mass ratio can further be converted into a more useful constraint on the dark matter mass fraction of the neutron star
\begin{equation}
\ratqm \equiv 1.22\times 10^{17} f_{DM} \left(\frac{g^2}{4\pi}\right)^{1/2}\left(\frac{100\text{ GeV}}{m_\chi}\right).
\end{equation}
The value of the self-interaction $g^2/4\pi$ is constrained primarily by astrophysical constraints on dark matter self-interactions, e.g. morphology of galactic halos. In particular, the ellipticity of large halos constrains $g^2/4\pi \lesssim 10^{-3}$ \cite{Petraki:2014uza}.  Saturating this bound, we see from Fig.~\ref{Fig:fconstraint} that for sub-TeV mass dark matter, gravitational waves can probe even the extreme dark matter mass fraction $f_{DM}\sim 10^{-15}$ in \cite{Kopp:2018jom}.

\vspace{-1mm}

\section{Discussion}
\label{sec:conclusions}

Current gravitational wave interferometers have been a remarkable success, and the observations of black-hole binary mergers  \cite{Abbott:2017gyy,Abbott:2017vtc,Abbott:2016nmj,Abbott:2016blz,Abbott:2017oio} and a neutron star binary merger \cite{TheLIGOScientific:2017qsa} have already place strong constraints on fundamental physics. The third generation of detectors will improve on LIGO sensitivity by up to two orders of magnitude, which provides ample cause for excitement at the prospect of further probing fundamental physics with gravitational waves from binary mergers.

In this work we have quantified these expectations, and have studied dark sector modifications to the gravitational waves emitted in binary inspirals. We have considered Yukawa corrections to the gravitational potential, and the associated dipole emission, as both arise in dark matter models with massive gauge bosons, and any modification of gravity that introduces a new scalar degree of freedom. We have explicitly computed the waveform, and performed a Fisher information matrix analysis to compute projected sensitivities of ten next generation gravitational wave detectors. 

The projected sensitivities to the Yukawa interaction coupling $\alpha$ and the dipole emission parameter $\gamma$ are shown in Figures \ref{fig:gammaplot} and \ref{fig:alphaplotNSNS}. The Einstein Telescope is found to be the most sensitive to such dark sector modifications, with sensitivity as good as $\mathcal{O}(10^{-5})$ and $\mathcal{O}(10^{-7})$ for $\alpha$ and $\gamma$ respectively. We project that constraints can be placed provided the Yukawa length scale $\lambda > \mathcal{O}(10)$ km, and they are optimal when $\lambda \sim 10^2 - 10^3$ km and $ \sim 10^{4}$ km  for $\alpha$ and $\gamma$ respectively.  The degree to which we can constraint these parameters is dependent on the signal-to-noise ratio of the gravitational wave detection.  Thus, parameters such as the masses of the binary constituents and the effective luminosity distance will play a significant role in the ability to constrain $\alpha,\gamma$.  Because the dark sector corrections considered here are not degenerate with higher PN corrections of GR, the spin parameters will not noticeably change the constraints.

\begin{figure}
\begin{center}
\includegraphics[width=.5 \textwidth]{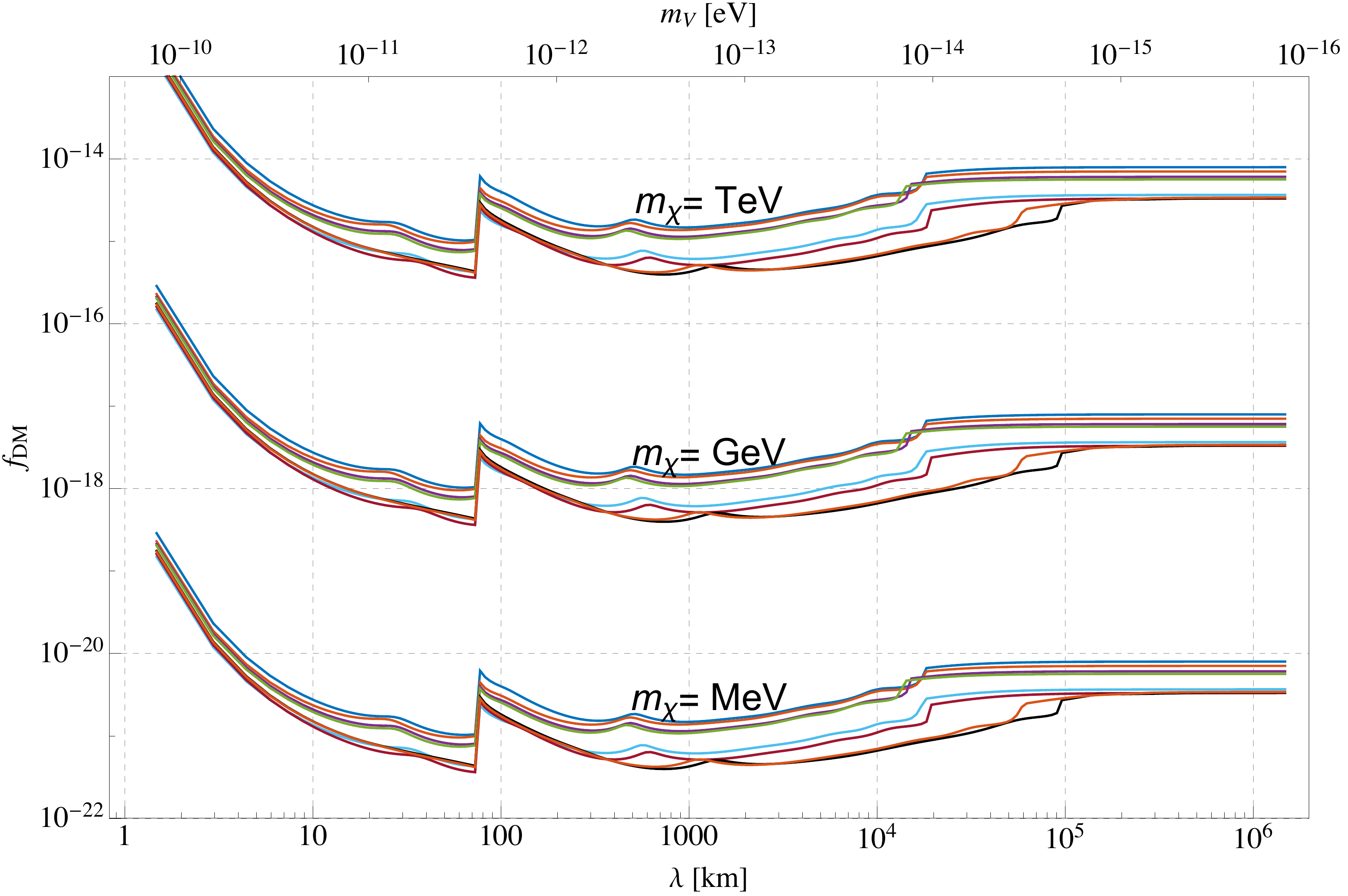}
\caption{Projected sensitivity to dark matter mass fraction from an NSNS binary merger, found from Eq.~(\ref{EQ:qmConst}), with $g^2/4 \pi = 10^{-3}$ and for varying mass $m_{\chi}$. Colors are as in previous figures.  At length scales below $\lambda ~ 70$ km, dipole radiation is not activated, and Eq.~(\ref{EQ:qmConst}) provides no constraint on the dark matter mass fraction.  One can provide optimistic constraints below this regime by assuming the mass fraction for the two neutron stars are comparable ($\gamma \ll \alpha$).}
\label{Fig:fconstraint}
\end{center}
\vspace{0cm}
\end{figure}

We emphasize that for a large range in $\lambda$, the Einstein Telescope gives the most stringent constraints for both $\alpha, \gamma$ parameters, due to the increased frequency range in the integration of the Fisher elements.  One may expect that the use of lower frequency detectors, such as LISA, may significantly improve these constraints.  But these space-based detectors will observe near-monochromatic binaries, so it is not clear whether these detectors will be effective at constraining dark sector modifications.

When written as a constraint on a specific dark matter model, we find these observations can detect even a minuscule amount of dark matter stored in neutron stars. For a GeV dark matter candidate with a gauge coupling $g^2/4\pi =10^{-3}$, the bound on the fraction of the NS mass in dark matter can easily be better than 1 part in $10^{15}$, as shown in Fig.~\ref{Fig:fconstraint}. More generally, the constraints on $\alpha$ and $\gamma$, shown in Figs.~\ref{fig:gammaplot} and \ref{fig:alphaplotNSNS}, probe dark photon masses in the range $m_v \lesssim 10^{-10}$ eV, with optimal constraints around $m_v \sim 10^{-12}$ eV.

We interpret these results as quantitative confirmation that gravitational wave astronomy is a powerful probe of fundamental physics. However, the work is not over, and there are indeed new directions for future work in every step on this analysis. In particular, one could improve upon theoretical estimates of the dark matter fraction of neutron stars,  extend the statistical analysis to include space-based detectors such as LISA (using extreme mass-ratio inspirals that include a neutron star component), and recompute the projected sensitivities by performing a full Markov-chain Monte-Carlo analysis. The last of these is a necessary step to properly quantify the degeneracy with astrophysical parameters, as well as the `fundamental theoretical bias' \cite{Yunes:2009ke} introduced by the use of GR waveforms and neglecting the modifications studied here. Each of these directions will be studied in future work.


\section*{Acknowledgements}

The authors thank Clare Burrage and Helvi Witek for insightful comments and suggestions. EM is supported in part by the National Science and Engineering Research Council of Canada via a PDF fellowship. N.~Y.~ acknowledges support from NSF CAREER grant
PHY-1250636 and NASA grants NNX16AB98G and 80NSSC17M0041.

\vspace{-.2cm}

\appendix

\section{Corrections to dipole radiation step function}
\label{App:DipoleRadCorrections}

We begin with the time-averaged power radiated through dipole emission of a vector or scalar source given by \cite{Krause:1994ar}:
\begin{align}
\langle \dot{E}_S \rangle &= \frac{1}{3}\eta^2 m^2 \omega^4 r^2 g_S(m_S,e)(\ratqm_1-\ratqm_2)^2,\\
\langle \dot{E}_V \rangle &= \frac{2}{3}\eta^2 m^2 \omega^4 r^2 g_V(m_V,e)(\ratqm_1-\ratqm_2)^2,
\end{align}
where $\ratqm_i$ is the charge-to-mass ratio of the compact object, and the $g_i$ functions are dependent on the eccentricity $e$ of the orbit and the mass of the additional degree of freedom.  Explicitly written,
\begin{widetext}
\begin{align}
g_S(m_S,e) &= \sum_{n=1}^\infty 2n^2\left[\mathcal{J}_n'^2(ne)+\left(\frac{1-e^2}{e^2}\right)\mathcal{J}_n^2(ne)\right]\times\left[1-\left(\frac{m_S}{n\omega}\right)^2\right]^{3/2},\\
g_V(m_V,e) &= \sum_{n=1}^\infty 2n^2\left[\mathcal{J}_n'^2(ne)+\left(\frac{1-e^2}{e^2}\right)\mathcal{J}_n^2(ne)\right]\times\left[1-\left(\frac{m_V}{n\omega}\right)^2\right]^{1/2}\left[1+\frac{1}{2}\left(\frac{m_V}{n\omega}\right)^2\right],
\end{align}
\end{widetext}
where $\mathcal{J}_n$ is the nth order Bessel function.  By taking the $e\rightarrow 0$ limit (circular orbits), we can use the identity
\begin{equation}
\lim_{e\rightarrow 0}\left[\mathcal{J}_n'^2(ne)+\left(\frac{1-e^2}{e^2}\right)\mathcal{J}_n^2(ne)\right] = \frac{1}{2}\delta_{n,1},
\end{equation}
to rewrite the time-averaged power radiated in the simple form
\begin{align}
\langle \dot{E}_S \rangle &= \frac{1}{3}\eta^2 m^2 \omega^4 r^2 (\ratqm_1-\ratqm_2)^2 \theta\left(\omega - m_S\right)\left(\frac{\omega^2-m_S^2}{\omega^2}\right)^{3/2},\label{EQ:DipS}\\
\langle \dot{E}_V \rangle &= \frac{2}{3}\eta^2 m^2 \omega^4 r^2 (\ratqm_1-\ratqm_2)^2 \theta\left(\omega - m_V\right)\nonumber\\
&\;\;\;\;\times\left(\frac{\omega^2-m_V^2}{\omega^2}\right)^{1/2}\left(\frac{\omega^2+m_V^2}{2\omega^2}\right).\label{EQ:DipV}
\end{align}
Note that if we ignore the ``corrections'' to the Heaviside step-function $\theta$ at high angular orbital frequencies ($\omega \gg m_{S,V}$), dipole radiation of a vector mode emits twice that of a scalar mode, but has the same functional form.

We now calculate the waveform including the dipole radiation term for either scalar or vector modes.  A more useful form will be as a ratio of $P_{GW}$:
\begin{align}
\frac{\langle \dot{E}_S \rangle}{P_{GW}} &= \frac{5(\ratqm_1-\ratqm_2)^2}{96m^{2/3}\omega^{2/3}}\left(\frac{\lambda_S^2\omega^2-1}{\lambda_S^2\omega^2}\right)^{3/2},\\
\frac{\langle \dot{E}_V \rangle}{P_{GW}} &= \frac{5(\ratqm_1-\ratqm_2)^2}{48m^{2/3}\omega^{2/3}}\left(\frac{\lambda_V^2\omega^2-1}{\lambda_V^2\omega^2}\right)^{1/2}\left(\frac{2\lambda_V^2\omega^2+1}{2\lambda_V^2\omega^2}\right),
\end{align}
where $\lambda_i = m_i^{-1}$ is the length scale associated with the additional scalar or vector degree of freedom.  The introduction of dipole radiation will manifest as an additional factor in the equation for $\dot{\omega}$.  In particular, we assume $(\ratqm_1 - \ratqm_2)^2 \ll 1$ so that dipole radiation is a small correction to the usual gravitational radiation.  Then,
\begin{equation}
\dot{\omega}^{-1} = \frac{5}{96}{\cal{M}}^{-5/3}\omega^{-11/3}\left(1-\frac{\langle \dot{E}_i \rangle}{P_{GW}}\right).
\end{equation}

In order to calculate the phase of the gravitational waveform, we must integrate the function
\begin{align}
2\omega t - 2\phi &= 2\int^{\omega}\frac{\omega - \omega'}{\dot{\omega}'}d\omega'\label{EQ:IntDef}\\
&= \frac{5}{48}{\cal{M}}^{-5/3}\int^\omega (\omega-\omega')\omega'^{-11/3}\left(1-\frac{\langle \dot{E}_i \rangle}{P_{GW}}\right) d\omega'.\nonumber
\end{align}
Including the corrections to the Heaviside step function, the dipole term results can be integrated in terms of hypergeometric functions.  However, we wish to find a power series expansion for the integral.  We expand each function as
\begin{equation}
\frac{\langle \dot{E}_i \rangle}{P_{GW}} = C_i\theta\left(\lambda\omega - 1\right)\sum_{n=0}^{\infty}(-1)^n a_i(n)\left(\lambda_i\omega\right)^{-2n},
\end{equation}
where $i = S,V$ denote the type of dipole radiation, and
\begin{align}
C_s &= \frac{5(\ratqm_1-\ratqm_2)^2}{96 m^{2/3}},\\
C_v &= \frac{5(\ratqm_1-\ratqm_2)^2}{48 m^{2/3}},\\
a_s(n) &= \frac{3\sqrt{\pi}}{4\Gamma[\frac{5}{2}-n]\Gamma[n+1]},\\
a_v(n) &= \frac{3\sqrt{\pi}(1-n)}{4\Gamma[\frac{5}{2}-n]\Gamma[n+1]}.
\end{align}
The integral in Eq.~(\ref{EQ:IntDef}) is then evaluated as
\begin{widetext}
\begin{align}
2\omega t - 2\phi = 2\omega &\left[t_0 - \delta t_0 \theta(\lambda_i\omega - 1)\right] - 2\left[\phi_0 - \delta \phi_0 \theta(\lambda_i\omega - 1)\right]\nonumber\\
&+ \frac{3}{128}({\cal{M}} \omega)^{-5/3}\left[1-20 C_i\omega^{-2/3}\theta(\lambda_i\omega -1)\sum_{n=0}^\infty \frac{(-1)^n a_i(n)}{(3n+5)(6n+7)}\left(\lambda_i\omega\right)^{-2n}\right].
\end{align}
\end{widetext}
We note, the corrections to the coalescence time $t_0$ and inspiral phase $\phi_0$ include nontrivial frequency dependence through the step function.  In principle, these additional step function corrections can be important for the matched filter process.  However, these corrections enter at 2.5PN and 4PN order for the phase and coalescence time, respectively, and should be small for most observations.

Finally, the waveform is written as
\begin{widetext}
\begin{align}
\tilde{h}(f) &= -\left(\frac{5\pi}{24}\right)^{1/2}\frac{{\cal{M}}^2}{D_\text{eff}}(\pi {\cal{M}} f)^{-7/6}\left[1-\frac{1}{2}C_i(\pi f)^{-2/3}\theta(\pi\lambda_i f- 1)\sum_{n=0}^\infty (-1)^n a_i(n)\left(\pi \lambda_i f\right)^{-2n}\right]e^{-i\Psi},\\
\Psi &= 2\omega \left[t_0 - \delta t_0 \theta(\pi\lambda_i f - 1)\right] - 2\left[\phi_0 - \delta \phi_0 \theta(\pi\lambda_i f - 1)\right]-\frac{\pi}{4}\nonumber\\
&+ \frac{3}{128}(\pi {\cal{M}} f)^{-5/3}\left[1-20 C_i(\pi f)^{-2/3}\theta(\pi\lambda_i f - 1)\sum_{n=0}^\infty \frac{(-1)^n a_i(n)}{(3n+5)(6n+7)}\left(\pi\lambda_i f\right)^{-2n}\right].
\end{align}
\end{widetext}
Due to the step function, $1\leq \pi\lambda_i f$, the infinite sum converges (to the same hypergeometric functions stated before) for both scalar and vector modes.  In the case of vector mode dipole radiation, $a_v(1) = 0$, hence the first correction to the step function occurs at second order, $(\pi\lambda_i f)^{-4}$.  Then, the $-1$PN correction to the waveform from the step-function dipole term is modified by a small $-7$PN correction for vector mode radiation or a $-4$PN correction for scalar mode radiation (small in the sense that $(\pi \lambda_i f)^{-2n}\leq 1$).

\section{Analytic Fits to Projected Detector Sensitivities}
\label{appFits}
We perform analytic fits to the tabulated noise curve for each detector. The functional form we use is
\be
\frac{1}{2}\log S_n (f) = \sum_{i=1} ^9 p_i \left( \frac{x - p_{10}}{p_{11}}\right)^{9-i} + \frac{p_{12}}{\sqrt{(x-p_{13})^{2} + p_{14}^2}}
\ee
where $x \equiv \log f$.  The final term is only included when the detector obtains a large resonance at small frequencies near $f_\text{low-cut}$.  This resonance does not occur in CE2 (narrow and wide) and the Einstein Telescope, hence we set $p_{12} = 0$ for these four fit functions.  While the shift and rescaling parameters $p_{10},p_{11}$ are redundant in this expansion, $p_{10}$ will manifest as a ``characteristic'' frequency, similar to previous work \cite{Will:1994fb}. The fitting parameters are given explicitly in Table~\ref{tab:fits}. 
 
\begin{table*}[t!]
\begin{tabularx}{1\textwidth}{|Y | Y | Y | Y | Y | Y | Y | Y | Y | Y | Y|}
\hline \hline
	  &  ALIGO  &  A+  & A++ &  VRT  &  Voyager  &  CE1  &  CE2w  &  CE2n  &  ET-B &  ET-D \\ \hline
	$p_1$  & $ 0.1136 $ & $ 9.235\times 10^{-2} $ & $ 7.22\times 10^{-2} $ & $ -1.379\times 10^{-2} $ & $ -1.149\times 10^{-5} $ & $ 0.2076 $ & $ 0.1111 $ & $ 0.1021 $ & $ 0.1132 $ & $ 0.3328 $ \\ \hline
	$p_2$  & $ -3.296\times 10^{-2} $ & $ -5.346\times 10^{-2} $ & $ -1.006\times 10^{-4} $ & $ -4.705\times 10^{-2} $ & $ 4.806\times 10^{-4} $ & $ -0.2575$ & $ -0.1297 $ & $ -0.1333 $ & $ -0.194 $ & $ -0.3963$ \\ \hline
	$p_3$  & $ -0.5891 $ & $ -0.4692 $ & $ -0.3896 $ & $ 0.2288 $ & $ -7.609\times 10^{-3} $ & $ -0.8017 $ & $ -0.4107$ & $ -0.3544 $ & $ -0.4398 $ & $ -1.122 $ \\ \hline
	$p_4$  & $ -4.839\times 10^{-3} $ & $ 8.376\times 10^{-2} $ & $ -0.2102 $ & $ -0.1593 $ & $ 4.603 \times 10^{-2} $ & $ 0.7220 $ & $ 0.2321 $ & $ 0.3786 $ & $ 0.9275 $ & $ 0.6489 $ \\ \hline
	$p_5$  & $ 1.141 $ & $ 0.9587 $ & $ 0.9982 $ & $ -0.2686$ & $ 0.1295 $ & $ 1.0520 $ & $ 0.6513 $ & $ 0.4214 $ & $ 8.287 \times 10^{-2} $ & $ 1.568 $ \\ \hline
	$p_6$ & $ -6.951\times 10^{-2} $ & $ -0.1183 $ & $ 0.2691 $ & $ 0.5208 $ & $ -3.569 $ & $ -0.4328 $ & $ 3.022\times 10^{-2} $ & $ -0.6310 $ & $ -1.428 $ & $ 0.2107 $ \\ \hline
	$p_7 $ & $ 0.2893 $ & $ 0.3372 $ & $ 7.438\times 10^{-2} $ & $ 0.6479 $ & $ 21.42 $ & $ 0.2003 $ & $ 0.2179 $ & $ 0.8405 $ & $ 2.767 $ & $ 0.3309 $ \\ \hline
	$p_8 $ & $ 7.484\times 10^{-2} $ & $ -0.3580 $ & $ -0.4025 $ & $ -0.3940 $ & $ -57.99 $ & $ 0.2033 $ & $ 0.2460 $ & $ 1.238 $ & $ -0.5816$ & $ -0.9473 $ \\ \hline
	$p_9 $ & $ -54 $ & $ -54.6 $ & $ -54.79 $ & $ -54.96 $ & $ 6.253 $ & $ -56.66 $ & $ -56.98 $ & $ -57.2 $ & $ -56.63 $ & $ -56.12 $ \\ \hline
	$p_{10} $ & $ 5.536 $ & $ 5.536 $ & $ 5.536 $ & $ 5.536 $ & $ 0 $ & $ 5.445 $ & $ 5.41 $ & $ 5.41 $ & $ 4.605 $ & $ 4.605 $ \\ \hline
	$p_{11} $ & $ 2.144 $ & $ 2.144 $ & $ 2.144 $ & $ 2.144 $ & $ 1 $ & $ 2.174$ & $ 2.196 $ & $ 2.196 $ & $ 2.66 $ & $ 2.66 $ \\ \hline
	$p_{12} $ & $ 4.003\times 10^{-2} $ & $ 4.395\times 10^{-2} $ & $ 4.771\times 10^{-2} $ & $ 1.762\times 10^{-2} $ & $ 3.153\times 10^{-2} $ & $ 1.429\times 10^{-2} $ & $ 0 $ & $ 0 $ & $ 0 $ & $ 0 $ \\ \hline
	$p_{13} $ & $ 2.275 $ & $ 2.275 $ & $ 2.275 $ & $ 1.821 $ & $ 1.921 $ & $ 1.633 $ & $ - $ & $ - $ & $ - $ & $ - $ \\ \hline
	$p_{14} $ & $ 9.254\times 10^{-3} $ & $ 1.020\times 10^{-2} $ & $ 1.108\times 10^{-2} $ & $ 2.954\times 10^{-3} $ & $ 7.960\times 10^{-3} $ & $ 3.879\times 10^{-3} $ & $ - $ & $ - $ & $ - $ & $ - $ \\ \hline \hline
\end{tabularx}
\caption{\label{tab:fits} Fitting parameters for noise curves.}
\end{table*}


\begin{thebibliography}{99}

\bibitem{Abbott:2017oio} 
  B.~P.~Abbott {\it et al.} [LIGO Scientific and Virgo Collaborations],
  \emph{GW170814: A Three-Detector Observation of Gravitational Waves from a Binary Black Hole Coalescence},
  Phys.\ Rev.\ Lett.\  {\bf 119}, no. 14, 141101 (2017)
  [arXiv:1709.09660 [gr-qc]].

\bibitem{Abbott:2017gyy} 
  B.~. P.~.Abbott {\it et al.} [LIGO Scientific and Virgo Collaborations],
  \emph{GW170608: Observation of a 19-solar-mass Binary Black Hole Coalescence},
  Astrophys.\ J.\  {\bf 851}, no. 2, L35 (2017)
  [arXiv:1711.05578 [astro-ph.HE]].

\bibitem{Abbott:2017vtc} 
  B.~P.~Abbott {\it et al.} [LIGO Scientific and VIRGO Collaborations],
  \emph{GW170104: Observation of a 50-Solar-Mass Binary Black Hole Coalescence at Redshift 0.2},
  Phys.\ Rev.\ Lett.\  {\bf 118}, no. 22, 221101 (2017)
  [arXiv:1706.01812 [gr-qc]].
  
\bibitem{Abbott:2016nmj} 
  B.~P.~Abbott {\it et al.} [LIGO Scientific and Virgo Collaborations],
  \emph{GW151226: Observation of Gravitational Waves from a 22-Solar-Mass Binary Black Hole Coalescence},
  Phys.\ Rev.\ Lett.\  {\bf 116}, no. 24, 241103 (2016)
  [arXiv:1606.04855 [gr-qc]].
  
\bibitem{Abbott:2016blz} 
  B.~P.~Abbott {\it et al.} [LIGO Scientific and Virgo Collaborations],
  \emph{Observation of Gravitational Waves from a Binary Black Hole Merger},
  Phys.\ Rev.\ Lett.\  {\bf 116}, no. 6, 061102 (2016)
  [arXiv:1602.03837 [gr-qc]].
  
\bibitem{TheLIGOScientific:2017qsa} 
  B.~P.~Abbott {\it et al.} [LIGO Scientific and Virgo Collaborations],
  \emph{GW170817: Observation of Gravitational Waves from a Binary Neutron Star Inspiral},
  Phys.\ Rev.\ Lett.\  {\bf 119}, no. 16, 161101 (2017)
  [arXiv:1710.05832 [gr-qc]].
  
\bibitem{Chamberlain:2017fjl} 
  K.~Chamberlain and N.~Yunes,
  \emph{Theoretical Physics Implications of Gravitational Wave Observation with Future Detectors},
  Phys.\ Rev.\ D {\bf 96}, no. 8, 084039 (2017)
  [arXiv:1704.08268 [gr-qc]].
  
\bibitem{Yunes:2016jcc} 
  N.~Yunes, K.~Yagi and F.~Pretorius,
  \emph{Theoretical Physics Implications of the Binary Black-Hole Mergers GW150914 and GW151226},
  Phys.\ Rev.\ D {\bf 94}, no. 8, 084002 (2016)
  [arXiv:1603.08955 [gr-qc]].
  
\bibitem{Ezquiaga:2017ekz} 
  J.~M.~Ezquiaga and M.~Zumalacarregui,
  \emph{Dark Energy After GW170817: Dead Ends and the Road Ahead},
  Phys.\ Rev.\ Lett.\  {\bf 119}, no. 25, 251304 (2017)
  [arXiv:1710.05901 [astro-ph.CO]].
  
\bibitem{Creminelli:2017sry} 
  P.~Creminelli and F.~Vernizzi,
  \emph{Dark Energy after GW170817 and GRB170817A},
  Phys.\ Rev.\ Lett.\  {\bf 119}, no. 25, 251302 (2017)
  [arXiv:1710.05877 [astro-ph.CO]].
  
\bibitem{Sakstein:2017xjx} 
  J.~Sakstein and B.~Jain,
  \emph{Implications of the Neutron Star Merger GW170817 for Cosmological Scalar-Tensor Theories},
  Phys.\ Rev.\ Lett.\  {\bf 119}, no. 25, 251303 (2017)
  [arXiv:1710.05893 [astro-ph.CO]].
  
\bibitem{ay1} 
  S.~Alexander and N.~Yunes,
  \emph{A New PPN parameter to test Chern-Simons gravity},
  Phys.\ Rev.\ Lett.\  {\bf 99}, 241101 (2007)
  [hep-th/0703265].
  
\bibitem{ay2} 
  S.~Alexander, L.~S.~Finn and N.~Yunes,
  \emph{A Gravitational-wave probe of effective quantum gravity},
  Phys.\ Rev.\ D {\bf 78}, 066005 (2008)
  [arXiv:0712.2542 [gr-qc]].

\bibitem{ay3} 
  S.~H.~Alexander and N.~Yunes,
  \emph{Gravitational wave probes of parity violation in compact binary coalescences},
  Phys.\ Rev.\ D {\bf 97}, no. 6, 064033 (2018)
  [arXiv:1712.01853 [gr-qc]].

\bibitem{Visinelli:2017bny} 
  L.~Visinelli, N.~Bolis and S.~Vagnozzi,
  \emph{Brane-world extra dimensions in light of GW170817},
  Phys.\ Rev.\ D {\bf 97}, no. 6, 064039 (2018)
  [arXiv:1711.06628 [gr-qc]].

\bibitem{Pardo:2018ipy} 
  K.~Pardo, M.~Fishbach, D.~E.~Holz and D.~N.~Spergel,
  \emph{Limits on the number of spacetime dimensions from GW170817},
  JCAP {\bf 1807}, no. 07, 048 (2018)
  [arXiv:1801.08160 [gr-qc]].
  
  
  
\bibitem{Amendola:2017ovw} 
  L.~Amendola, I.~Sawicki, M.~Kunz and I.~D.~Saltas,
  \emph{Direct detection of gravitational waves can measure the time variation of the Planck mass},
  arXiv:1712.08623 [astro-ph.CO].
  
\bibitem{Yagi:2013qpa} 
  K.~Yagi, D.~Blas, N.~Yunes and E.~Barausse,
  \emph{Strong Binary Pulsar Constraints on Lorentz Violation in Gravity},
  Phys.\ Rev.\ Lett.\  {\bf 112}, no. 16, 161101 (2014)
  [arXiv:1307.6219 [gr-qc]].
  
\bibitem{Yagi:2013ava} 
  K.~Yagi, D.~Blas, E.~Barausse and N.~Yunes,
  \emph{Constraints on Einstein-Aether theory and Horava gravity from binary pulsar observations},
  Phys.\ Rev.\ D {\bf 89}, no. 8, 084067 (2014)
  Erratum: [Phys.\ Rev.\ D {\bf 90}, no. 6, 069902 (2014)]
  Erratum: [Phys.\ Rev.\ D {\bf 90}, no. 6, 069901 (2014)]
  [arXiv:1311.7144 [gr-qc]].

\bibitem{Yunes:2011aa} 
  N.~Yunes, P.~Pani and V.~Cardoso,
  \emph{Gravitational Waves from Quasicircular Extreme Mass-Ratio Inspirals as Probes of Scalar-Tensor Theories},
  Phys.\ Rev.\ D {\bf 85}, 102003 (2012)
  [arXiv:1112.3351 [gr-qc]].
  
\bibitem{Damour:1993hw} 
  T.~Damour and G.~Esposito-Farese,
  \emph{Nonperturbative strong field effects in tensor - scalar theories of gravitation},
  Phys.\ Rev.\ Lett.\  {\bf 70}, 2220 (1993).
  
\bibitem{Damour:1996ke} 
  T.~Damour and G.~Esposito-Farese,
  \emph{Tensor - scalar gravity and binary pulsar experiments},
  Phys.\ Rev.\ D {\bf 54}, 1474 (1996)
  [gr-qc/9602056].
  
\bibitem{Mirshekari:2013vb} 
  S.~Mirshekari and C.~M.~Will,
  \emph{Compact binary systems in scalar-tensor gravity: Equations of motion to 2.5 post-Newtonian order},
  Phys.\ Rev.\ D {\bf 87}, no. 8, 084070 (2013)
  [arXiv:1301.4680 [gr-qc]].
  
\bibitem{Lombriser:2016yzn} 
  L.~Lombriser and N.~A.~Lima,
  \emph{Challenges to Self-Acceleration in Modified Gravity from Gravitational Waves and Large-Scale Structure},
  Phys.\ Lett.\ B {\bf 765}, 382 (2017)
  [arXiv:1602.07670 [astro-ph.CO]].
     
\bibitem{Aghanim:2018eyx} 
  N.~Aghanim {\it et al.} [Planck Collaboration],
  \emph{Planck 2018 results. VI. Cosmological parameters},
  arXiv:1807.06209 [astro-ph.CO].
  
\bibitem{Rubin:1970zza} 
  V.~C.~Rubin and W.~K.~Ford, Jr.,
  \emph{Rotation of the Andromeda Nebula from a Spectroscopic Survey of Emission Regions},
  Astrophys.\ J.\  {\bf 159}, 379 (1970).
  
\bibitem{Rubin:1980zd} 
  V.~C.~Rubin, N.~Thonnard and W.~K.~Ford, Jr.,
  \emph{Rotational properties of 21 SC galaxies with a large range of luminosities and radii, from NGC 4605 /R = 4kpc/ to UGC 2885 /R = 122 kpc},
  Astrophys.\ J.\  {\bf 238}, 471 (1980).

\bibitem{Clowe:2006eq} 
D.~Clowe, A.~Gonzalez and M.~Markevitch,
  \emph{Weak lensing mass reconstruction of the interacting cluster 1E0657-558: Direct evidence for the existence of dark matter},
  Astrophys.\ J.\  {\bf 604}, 596 (2004)
  [astro-ph/0312273].
  
  D.~Clowe, M.~Bradac, A.~H.~Gonzalez, M.~Markevitch, S.~W.~Randall, C.~Jones and D.~Zaritsky,
  \emph{A direct empirical proof of the existence of dark matter},
  Astrophys.\ J.\  {\bf 648}, L109 (2006)
  [astro-ph/0608407].

\bibitem{Alexander:2018fjp} 
  S.~Alexander, E.~McDonough and D.~N.~Spergel,
  \emph{Chiral Gravitational Waves and Baryon Superfluid Dark Matter},
  JCAP {\bf 1805}, no. 05, 003 (2018)
  [arXiv:1801.07255 [hep-th]].
  
\bibitem{Guzzetti:2016mkm} 
  M.~C.~Guzzetti, N.~Bartolo, M.~Liguori and S.~Matarrese,
  \emph{Gravitational waves from inflation},
  Riv.\ Nuovo Cim.\  {\bf 39}, no. 9, 399 (2016)
  [arXiv:1605.01615 [astro-ph.CO]].
  
\bibitem{Bird:2016dcv} 
  S.~Bird, I.~Cholis, J.~B.~Munoz, Y.~Ali-Ha\"{i}moud, M.~Kamionkowski, E.~D.~Kovetz, A.~Raccanelli and A.~G.~Riess,
  \emph{Did LIGO detect dark matter?},
  Phys.\ Rev.\ Lett.\  {\bf 116}, no. 20, 201301 (2016)
  [arXiv:1603.00464 [astro-ph.CO]].

\bibitem{Goldman:1989nd} 
  I.~Goldman and S.~Nussinov,
  \emph{Weakly Interacting Massive Particles and Neutron Stars},
  Phys.\ Rev.\ D {\bf 40}, 3221 (1989).
  
\bibitem{Kouvaris:2007ay} 
  C.~Kouvaris,
  \emph{WIMP Annihilation and Cooling of Neutron Stars},
  Phys.\ Rev.\ D {\bf 77}, 023006 (2008)
  [arXiv:0708.2362 [astro-ph]].

\bibitem{Kouvaris:2010vv} 
  C.~Kouvaris and P.~Tinyakov,
  \emph{Can Neutron stars constrain Dark Matter?},
  Phys.\ Rev.\ D {\bf 82}, 063531 (2010)
  [arXiv:1004.0586 [astro-ph.GA]].

\bibitem{deLavallaz:2010wp} 
  A.~de Lavallaz and M.~Fairbairn,
  \emph{Neutron Stars as Dark Matter Probes},
  Phys.\ Rev.\ D {\bf 81}, 123521 (2010)
  [arXiv:1004.0629 [astro-ph.GA]].
  
\bibitem{1012.2039} 
  C.~Kouvaris and P.~Tinyakov,
  \emph{Constraining Asymmetric Dark Matter through observations of compact stars},
  Phys.\ Rev.\ D {\bf 83}, 083512 (2011)
  [arXiv:1012.2039 [astro-ph.HE]].

\bibitem{1103.5472} 
  S.~D.~McDermott, H.~B.~Yu and K.~M.~Zurek,
  \emph{Constraints on Scalar Asymmetric Dark Matter from Black Hole Formation in Neutron Stars},
  Phys.\ Rev.\ D {\bf 85}, 023519 (2012)
  [arXiv:1103.5472 [hep-ph]].
  
\bibitem{1201.2400} 
  T.~G\"{u}ver, A.~E.~Erkoca, M.~Hall Reno and I.~Sarcevic,
  \emph{On the  capture of dark matter by  neutron stars},
  JCAP {\bf 1405}, 013 (2014)
  [arXiv:1201.2400 [hep-ph]].

\bibitem{1301.0036} 
  J.~Bramante, K.~Fukushima and J.~Kumar,
  \emph{Constraints on bosonic dark matter from observation of old neutron stars},
  Phys.\ Rev.\ D {\bf 87}, no. 5, 055012 (2013)
  [arXiv:1301.0036 [hep-ph]].
  
\bibitem{1301.6811} 
  N.~F.~Bell, A.~Melatos and K.~Petraki,
  \emph{Realistic neutron star constraints on bosonic asymmetric dark matter},
  Phys.\ Rev.\ D {\bf 87}, no. 12, 123507 (2013)
  [arXiv:1301.6811 [hep-ph]].

\bibitem{1310.3509} 
  J.~Bramante, K.~Fukushima, J.~Kumar and E.~Stopnitzky,
  \emph{Bounds on self-interacting fermion dark matter from observations of old neutron stars},
  Phys.\ Rev.\ D {\bf 89}, no. 1, 015010 (2014)
  [arXiv:1310.3509 [hep-ph]].

\bibitem{Bramante:2017xlb} 
  J.~Bramante, A.~Delgado and A.~Martin,
  \emph{Multiscatter stellar capture of dark matter},
  Phys.\ Rev.\ D {\bf 96}, no. 6, 063002 (2017)
  [arXiv:1703.04043 [hep-ph]].

\bibitem{Bramante:2017ulk} 
  J.~Bramante, T.~Linden and Y.~D.~Tsai,
  \emph{Searching for dark matter with neutron star mergers and quiet kilonovae},
  Phys.\ Rev.\ D {\bf 97}, no. 5, 055016 (2018)
  [arXiv:1706.00001 [hep-ph]].
  
\bibitem{Zheng:2014fya} 
  H.~Zheng, K.~J.~Sun and L.~W.~Chen,
  \emph{Old neutron stars as probes of isospin-violating dark matter},
  Astrophys.\ J.\  {\bf 800}, no. 2, 141 (2015)
  [arXiv:1408.2926 [nucl-th]].
  
\bibitem{Sandin:2008db} 
  F.~Sandin and P.~Ciarcelluti,
  \emph{Effects of mirror dark matter on neutron stars},
  Astropart.\ Phys.\  {\bf 32}, 278 (2009)
  [arXiv:0809.2942 [astro-ph]].
  
\bibitem{Raj:2017wrv} 
  N.~Raj, P.~Tanedo and H.~B.~Yu,
  \emph{Neutron stars at the dark matter direct detection frontier},
  Phys.\ Rev.\ D {\bf 97}, no. 4, 043006 (2018)
  [arXiv:1707.09442 [hep-ph]].
  
\bibitem{Cardoso:2016olt} 
  V.~Cardoso, C.~F.~B.~Macedo, P.~Pani and V.~Ferrari,
  \emph{Black holes and gravitational waves in models of minicharged dark matter},
  JCAP {\bf 1605}, no. 05, 054 (2016)
  [arXiv:1604.07845 [hep-ph]].
  
\bibitem{Sagunski:2017nzb} 
  L.~Sagunski, J.~Zhang, M.~C.~Johnson, L.~Lehner, M.~Sakellariadou, S.~L.~Liebling, C.~Palenzuela and D.~Neilsen,
  \emph{Neutron star mergers as a probe of modifications of general relativity with finite-range scalar forces},
  Phys.\ Rev.\ D {\bf 97}, no. 6, 064016 (2018)
  [arXiv:1709.06634 [gr-qc]].
  
\bibitem{Croon:2017zcu} 
  D.~Croon, A.~E.~Nelson, C.~Sun, D.~G.~E.~Walker and Z.~Z.~Xianyu,
  \emph{Hidden-Sector Spectroscopy with Gravitational Waves from Binary Neutron Stars},
  Astrophys.\ J.\  {\bf 858}, no. 1, L2 (2018)
  [arXiv:1711.02096 [hep-ph]].
  
\bibitem{Ellis:2017jgp} 
  J.~Ellis, A.~Hektor, G.~H\"utsi, K.~Kannike, L.~Marzola, M.~Raidal and V.~Vaskonen,
  \emph{Search for Dark Matter Effects on Gravitational Signals from Neutron Star Mergers},
  Phys.\ Lett.\ B {\bf 781}, 607 (2018)
  [arXiv:1710.05540 [astro-ph.CO]].
  
\bibitem{Kopp:2018jom} 
  J.~Kopp, R.~Laha, T.~Opferkuch and W.~Shepherd,
  \emph{Cuckoo's Eggs in Neutron Stars: Can LIGO Hear Chirps from the Dark Sector?},
  arXiv:1807.02527 [hep-ph].
  
\bibitem{Nelson:2018xtr} 
  A.~Nelson, S.~Reddy and D.~Zhou,
  \emph{Dark halos around neutron stars and gravitational waves},
  arXiv:1803.03266 [hep-ph].
  
\bibitem{Huang:2018pbu} 
  J.~Huang, M.~C.~Johnson, L.~Sagunski, M.~Sakellariadou and J.~Zhang,
  \emph{Prospects for axion searches with Advanced LIGO through binary mergers},
  arXiv:1807.02133 [hep-ph].
  
\bibitem{Petraki:2013wwa} 
  K.~Petraki and R.~R.~Volkas,
  \emph{Review of asymmetric dark matter},
  Int.\ J.\ Mod.\ Phys.\ A {\bf 28}, 1330028 (2013)
  [arXiv:1305.4939 [hep-ph]].
  
\bibitem{Petraki:2014uza} 
  K.~Petraki, L.~Pearce and A.~Kusenko,
  \emph{Self-interacting asymmetric dark matter coupled to a light massive dark photon},
  JCAP {\bf 1407}, 039 (2014)
  [arXiv:1403.1077 [hep-ph]].
  
  
\bibitem{Ellis:2018bkr} 
  J.~Ellis, G.~H\"utsi, K.~Kannike, L.~Marzola, M.~Raidal and V.~Vaskonen,
  \emph{Dark Matter Effects On Neutron Star Properties},
  Phys.\ Rev.\ D {\bf 97}, no. 12, 123007 (2018)
  [arXiv:1804.01418 [astro-ph.CO]].
  
\bibitem{Rezaei:2018cuk} 
  Z.~Rezaei,
  \emph{Double dark-matter admixed neutron star},
  arXiv:1807.01781 [astro-ph.HE].
  
\bibitem{Leung:2011zz} 
  S.~C.~Leung, M.~C.~Chu and L.~M.~Lin,
  \emph{Dark-matter admixed neutron stars},
  Phys.\ Rev.\ D {\bf 84}, 107301 (2011)
  [arXiv:1111.1787 [astro-ph.CO]].
  
\bibitem{Krause:1994ar} 
  D.~Krause, H.~T.~Kloor and E.~Fischbach,
  \emph{Multipole radiation from massive fields: Application to binary pulsar systems},
  Phys.\ Rev.\ D {\bf 49}, 6892 (1994).

\bibitem{Damour:1994ya} 
  T.~Damour and A.~M.~Polyakov,
  \emph{String theory and gravity},
  Gen.\ Rel.\ Grav.\  {\bf 26}, 1171 (1994)
  [gr-qc/9411069].

\bibitem{Bertotti:2003rm} 
  B.~Bertotti, L.~Iess and P.~Tortora,
  Nature {\bf 425}, 374 (2003).
  doi:10.1038/nature01997

\bibitem{Will:1994fb} 
  C.~M.~Will,
  \emph{Testing scalar - tensor gravity with gravitational wave observations of inspiraling compact binaries},
  Phys.\ Rev.\ D {\bf 50}, 6058 (1994)
  [gr-qc/9406022].
  
\bibitem{Alsing:2011er} 
  J.~Alsing, E.~Berti, C.~M.~Will and H.~Zaglauer,
  \emph{Gravitational radiation from compact binary systems in the massive Brans-Dicke theory of gravity},
  Phys.\ Rev.\ D {\bf 85}, 064041 (2012)
  [arXiv:1112.4903 [gr-qc]].
  
\bibitem{Berti:2018cxi} 
  E.~Berti, K.~Yagi and N.~Yunes,
  \emph{Extreme Gravity Tests with Gravitational Waves from Compact Binary Coalescences: (I) Inspiral-Merger},
  Gen.\ Rel.\ Grav.\  {\bf 50}, no. 4, 46 (2018)
  [arXiv:1801.03208 [gr-qc]].

\bibitem{Allen:2005fk} 
  B.~Allen, W.~G.~Anderson, P.~R.~Brady, D.~A.~Brown and J.~D.~E.~Creighton,
  \emph{FINDCHIRP: An Algorithm for detection of gravitational waves from inspiraling compact binaries},
  Phys.\ Rev.\ D {\bf 85}, 122006 (2012)
  [gr-qc/0509116].

\bibitem{Yunes:2009yz} 
  N.~Yunes, K.~G.~Arun, E.~Berti and C.~M.~Will,
  \emph{Post-Circular Expansion of Eccentric Binary Inspirals: Fourier-Domain Waveforms in the Stationary Phase Approximation},
  Phys.\ Rev.\ D {\bf 80}, no. 8, 084001 (2009)
  Erratum: [Phys.\ Rev.\ D {\bf 89}, no. 10, 109901 (2014)]
  [arXiv:0906.0313 [gr-qc]].
  
\bibitem{Porter:2015eha} 
  E.~K.~Porter and N.~J.~Cornish,
  \emph{Fisher versus Bayes: A comparison of parameter estimation techniques for massive black hole binaries to high redshifts with eLISA},
  Phys.\ Rev.\ D {\bf 91}, no. 10, 104001 (2015)
  [arXiv:1502.05735 [gr-qc]].
  
\bibitem{Vallisneri:2007ev} 
  M.~Vallisneri,
  \emph{Use and abuse of the Fisher information matrix in the assessment of gravitational-wave parameter-estimation prospects},
  Phys.\ Rev.\ D {\bf 77}, 042001 (2008)
  [gr-qc/0703086 [GR-QC]].
  
\bibitem{Ajith:2007kx} 
  P.~Ajith {\it et al.},
  \emph{A Template bank for gravitational waveforms from coalescing binary black holes. I. Non-spinning binaries},
  Phys.\ Rev.\ D {\bf 77}, 104017 (2008)
  Erratum: [Phys.\ Rev.\ D {\bf 79}, 129901 (2009)]
  [arXiv:0710.2335 [gr-qc]].
   
\bibitem{Abbott:2016bqf} 
  B.~P.~Abbott {\it et al.} [LIGO Scientific and Virgo Collaborations],
  \emph{The basic physics of the binary black hole merger GW150914},
  Annalen Phys.\  {\bf 529}, no. 1-2, 1600209 (2017)
  doi:10.1002/andp.201600209
  [arXiv:1608.01940 [gr-qc]].
  
\bibitem{Giudice:2016zpa} 
  G.~F.~Giudice, M.~McCullough and A.~Urbano,
  \emph{Hunting for Dark Particles with Gravitational Waves},
  JCAP {\bf 1610}, no. 10, 001 (2016)
  [arXiv:1605.01209 [hep-ph]].
  
\bibitem{Lehner:2016lxy} 
  L.~Lehner, S.~L.~Liebling, C.~Palenzuela, O.~L.~Caballero, E.~O'Connor, M.~Anderson and D.~Neilsen,
  \emph{Unequal mass binary neutron star mergers and multimessenger signals},
  Class.\ Quant.\ Grav.\  {\bf 33}, no. 18, 184002 (2016)
  [arXiv:1603.00501 [gr-qc]].
  
\bibitem{Mandel:2014tca} 
  I.~Mandel, C.~P.~L.~Berry, F.~Ohme, S.~Fairhurst and W.~M.~Farr,
  \emph{Parameter estimation on compact binary coalescences with abruptly terminating gravitational waveforms},
  Class.\ Quant.\ Grav.\  {\bf 31}, 155005 (2014)
  [arXiv:1404.2382 [gr-qc]].
  
\bibitem{Yagi:2013awa} 
  K.~Yagi and N.~Yunes,
  \emph{I-Love-Q Relations in Neutron Stars and their Applications to Astrophysics, Gravitational Waves and Fundamental Physics},
  Phys.\ Rev.\ D {\bf 88}, no. 2, 023009 (2013)
  [arXiv:1303.1528 [gr-qc]].
  
\bibitem{Aasi:2013wya} 
  B.~P.~Abbott {\it et al.} [KAGRA and LIGO Scientific and VIRGO Collaborations],
  \emph{Prospects for Observing and Localizing Gravitational-Wave Transients with Advanced LIGO, Advanced Virgo and KAGRA},
  Living Rev.\ Rel.\  {\bf 21}, 3 (2018)
  [Living Rev.\ Rel.\  {\bf 19}, 1 (2016)]
  [arXiv:1304.0670 [gr-qc]].
 
\bibitem{LIGOwhitepaper}
L. S. Collaboration, \emph{LIGO-T15TBI: Instrument Science White Paper}.

\bibitem{Adhikari:2013kya} 
  R.~X.~Adhikari,
  Rev.\ Mod.\ Phys.\  {\bf 86}, 121 (2014)
  doi:10.1103/RevModPhys.86.121
  [arXiv:1305.5188 [gr-qc]].
  
\bibitem{ET}
Available at \href{http://www.et-gw.eu/index.php/etsensitivities}{http://www.et-gw.eu/index.php/etsensitivities}

\bibitem{Hild:2009ns} 
  S.~Hild, S.~Chelkowski, A.~Freise, J.~Franc, N.~Morgado, R.~Flaminio and R.~DeSalvo,
  \emph{A Xylophone Configuration for a third Generation Gravitational Wave Detector},
  Class.\ Quant.\ Grav.\  {\bf 27}, 015003 (2010)
  [arXiv:0906.2655 [gr-qc]].
  
\bibitem{Cardoso:2018tly} 
  V.~Cardoso, \'O.~J.~C.~Dias, G.~S.~Hartnett, M.~Middleton, P.~Pani and J.~E.~Santos,
  \emph{Constraining the mass of dark photons and axion-like particles through black-hole superradiance},
  JCAP {\bf 1803}, no. 03, 043 (2018)
  [arXiv:1801.01420 [gr-qc]].
  
\bibitem{Evans:2016mbw} 
  B.~P.~Abbott {\it et al.} [LIGO Scientific Collaboration],
  \emph{Exploring the Sensitivity of Next Generation Gravitational Wave Detectors},
  Class.\ Quant.\ Grav.\  {\bf 34}, no. 4, 044001 (2017)
  [arXiv:1607.08697 [astro-ph.IM]].
  
\bibitem{Khan:2015jqa} 
  S.~Khan, S.~Husa, M.~Hannam, F.~Ohme, M.~P\"urrer, X.~Jim\'enez Forteza and A.~Boh\'e,
  \emph{Frequency-domain gravitational waves from nonprecessing black-hole binaries. II. A phenomenological model for the advanced detector era},
  Phys.\ Rev.\ D {\bf 93}, no. 4, 044007 (2016)
  [arXiv:1508.07253 [gr-qc]].
  
\bibitem{Bekenstein:1971hc} 
  J.~D.~Bekenstein,
  \emph{Nonexistence of baryon number for static black holes},
  Phys.\ Rev.\ D {\bf 5}, 1239 (1972);
 
  J.~D.~Bekenstein,
  \emph{Nonexistence of baryon number for black holes. ii},
  Phys.\ Rev.\ D {\bf 5}, 2403 (1972).
  
\bibitem{Battye:2018ssx} 
  R.~A.~Battye, F.~Pace and D.~Trinh,
  \emph{Gravitational wave constraints on dark sector models},
  Phys.\ Rev.\ D {\bf 98}, no. 2, 023504 (2018)
  doi:10.1103/PhysRevD.98.023504
  [arXiv:1802.09447 [astro-ph.CO]].
  
\bibitem{Yunes:2009ke} 
  N.~Yunes and F.~Pretorius,
  \emph{Fundamental Theoretical Bias in Gravitational Wave Astrophysics and the Parameterized Post-Einsteinian Framework},
  Phys.\ Rev.\ D {\bf 80}, 122003 (2009)
  [arXiv:0909.3328 [gr-qc]].

\end{thebibliography}
\end{document}